\documentstyle[aps,epsf]{revtex}

\newcommand{\beq}{\begin{equation}}
\newcommand{\eeq}{\end{equation}}
\newcommand{\bea}{\begin{eqnarray}}
\newcommand{\eea}{\end{eqnarray}}
\newcommand{\nn}{\nonumber}
\newcommand{\junk}[1]{}

\def\gmf{\gamma _{5}}

\def\la{\langle }
\def\ra{ \rangle }

\def\lo{\langle 0 |}
\def\ro{ | 0 \rangle }

\newcommand{\q}{{\bf q}}
\newcommand{\C}{{\cal C}}
\newcommand{\D}{{\cal D}}

\newcommand{\Z}{{\rm Z\!\!Z}}
\newcommand{\MM}{{\cal M}}
\newcommand{\X}{{\bf X}}
\newcommand{\e}{{\rm e}}

 \tiny \normalsize

\begin{document}

 \hsize\textwidth\columnwidth\hsize\csname@twocolumnfalse\endcsname
\draft

\title{Monopoles and Coulomb Gas Representation of  the QCD
Effective   Lagrangian}
\author{Sebastian Jaimungal and Ariel R. Zhitnitsky}
\address{Department of Physics and Astronomy, University of British Columbia,
Vancouver, BC V6T 1Z1, Canada}
\date{\today}
\maketitle

\begin{abstract}
A novel Coulomb gas (CG) description of low energy $QCD_4(N_c)$ is
constructed.  The construction is based on the dual transformation of
the $QCD$ effective Lagrangian.  By considering a large gauge
transformation, the charges of this statistical system are identified
with magnetic monopoles which carry fractional charges of strength
$1/N_c$.  Furthermore, the creation operator which inserts the
magnetic charge in the CG picture is explicitly constructed and
demonstrated to have a non-zero vacuum expectation value, indicating
that confinement does occur. The Wilson loop operator as well as the
creation operator for the domain wall in CG representation is also
constructed.  Additional support for the D-brane picture suggested by
Witten is also found. Lastly, the relation of the CG picture with the
instanton-quarks is also discussed.
\end{abstract}
\pacs{PACS codes:}
\vspace{5mm}

\section{Introduction}

Color confinement, spontaneous breaking of chiral symmetry, the $U(1)$
problem, $\theta$ dependence, and the classification of vacuum states are
some of the most interesting questions in $QCD$. Unfortunately, the
progress in our understanding of them is extremely slow. At the end of
the 1970s A. M. Polyakov \cite{Po77} demonstrated color confinement in
$QED_3$; this was the first example in which nontrivial dynamics was a
key ingredient in the solution.  Soon after, 't Hooft and Mandelstam
\cite{Hooft} suggested a qualitative picture of how confinement could
occur in $QCD_4$.  The key point, the 't Hooft - Mandelstam approach,
is the assumption that dynamical monopoles exist and Bose
condense. Many papers have been written on this subject since the
original formulation \cite{Hooft}; however, the main questions, such
as, ``What are these monopoles?''; ``How do they appear in the gauge
theories without Higgs fields?''; ``How do they interact?'', were
still not understood (for a recent review see \cite{Hooft1}). Almost
20 years passed before the next important piece of the puzzle was
solved \cite{SeWi94}. Seiberg and Witten demonstrated that confinement
occurs in SUSY $QCD_4$ due to the condensation of monopoles much along
the lines suggested many years ago by 't Hooft and
Mandelstam. Furthermore, condensation of dyons together with oblique
confinement for nonzero vacuum angle, $\theta$, was also discovered in
SUSY models (a phenomenon which was also argued to take place in
ordinary $QCD$; see \cite{Hooft1}). In addition to forming concrete
realizations of earlier ideas, the recent progress in SUSY models has
introduced many new phenomena, such as the existence of domain
walls\cite{Shifman1} which connect two distinct $\theta$-vacua. New
insights into confinement was also recently given by Witten
\cite{Wi98}, in which he argued that domain walls connecting two vacua
labeled by $k$ and $k+1$ behave similarly to $D$-branes on which the
$SQCD$ strings can end. It is tantalizing to suggest that such
phenomenon also take place in $QCD$; indeed, in this paper it is
argued that this in fact does occur.

With such a wealth of information now available for SUSY gauge
theories, it is interesting to ask how one can apply that knowledge to
QCD. In a recent review Shifman \cite{Shifman1} addressed certain
aspects of this question and we refer to that paper for further
references and recent development. Thus far, most works have
approached such questions by starting with a supersymmetric theory
where some particular dynamical aspect of interest is already known;
then introducing explicit supersymmetry breaking terms by giving the
gluinos or squarks mass; finally, in the limit of infinite mass one
recovers a non-supersymmetric theory with decoupled gluinos and
squarks. Unfortunately, any calculations in this limit are beyond
reach. Consequently, theories with a small SUSY breaking mass term are
usually studied and a qualitative picture of what happens in QCD (and
other non-supersymmetric theories) is extracted. This procedure has
lead to success in the identification of the complicated vacuum structure
for such theories: there are $N$ inequivalent vacua for each given
$\theta$ \cite{Shifman1}. These vacua are distinguished by the phase
of the chiral condensate and there are domain walls which interpolate
between them. It is believed that in QCD such a similar structure
exists. 

Rather than beginning with the SUSY model and then breaking the
symmetry, our analysis begins with non-supersymmetric $QCD$. For the
purposes of the paper it will be assumed that the theory is in the
confining phase and that chiral symmetry is broken. The question to be
addressed is then: ``Does the picture which emerges in SUSY models
reflect what happens in QCD?''. We will demonstrate that the answer is
indeed {\it yes}. This strategy clearly does not prove confinement,
nor does it elaborate on the microscopic mechanism responsible for it,
however, it will, however, demonstrate that the assumption of
confinement implies the condensation of massless monopoles and vice
versa.  Furthermore, the role that the $\theta$ parameter plays in the
condensation of dyons with non-zero electric charge will be
illuminated. In addition, this strategy will allow us to demonstrate
that the states in QCD for each given $\theta$-angle is indeed
classified by the integer number $k$ labeling the phase of the dyon
condensate. Finally, domain walls which interpolate between these
different phases will be shown to have wall tension $\sim{N}$, and it
is consistent with their interpretation as $D$-branes rather than as
QCD string solitons. In addition, a relationship between the dual
representation of low energy QCD, in terms of the Coulomb Gas of
monopoles, and instantons will be discussed. This leads us to {\it
conjecture} that at large distances our particles with fractional
magnetic charges, from the CG representation, are in fact the
instanton-quarks suspected long ago \cite{Belavin}.

Before going into a detailed analysis of the effect Lagrangian the
starting point of our analysis will now be elaborated on.  Experience
in SUSY models demonstrates that the effective Lagrangian approach and
duality transformations are very effective tools in the analysis of
the large distance dynamics in the strong coupling regime (see
e.g. reviews \cite{review}).  These tools will be adopted for the
study of QCD. The first key element in the analysis is the effective
Lagrangian approach. As is well known, there are two different
definitions of an effective Lagrangian in quantum field theory. One of
them is the Wilsonian effective Lagrangian describing the low energy
dynamics of the lightest particles in the theory. In QCD, this is
implemented by effective chiral Lagrangians for the pseudoscalar
mesons.  Another type of the effective Lagrangian (potential) is
defined as the Legendre transform of the generating functional for
connected Greens functions. This object is useful in addressing
questions about the vacuum properties of the theory in terms of vacuum
expectation values (VEV's) of composite operators, as they minimize
the effective action.  Such an approach is well suited for studying
the dependence of the vacuum state on external parameters, such as the
light quark masses or the vacuum angle $\theta$. However, it is not
useful for studying, for instance, $S$-matrix elements because the
kinetic term cannot be recovered in such an approach. The utility of
such an approach to gauge theories has been recognized long ago for
supersymmetric models, where the anomalous effective potential has
been found for both the pure gauge case \cite{VY} and $SQCD$
\cite{TVY}. Properties of the vacuum structure in SUSY models were
correctly understood only after analyzing this kind of effective
potential. The second key point of our approach is the observation
that such an effective potential in QCD (which is a collection of
self-interacting Sine-Gordon (SG) fields) has a dual representation in
terms of a Coulomb Gas of several species of charges, which will be
described in detail in Section 3.

Since our description starts from color-singlet fields (the phases of
the chiral condensate), one might think that no information about
confinement, which is clearly a color phenomena, can be extracted from
such an analysis. However, contrary to ones instinct, it is possible
to obtain color information due to the existence of the free parameter
$\theta$, which plays the role of the messenger between colorless and
colorful objects.  Indeed, the dependence of the physical observables
as a function of $\theta$ can be studied exclusively in terms of the
colorless degrees of freedom by using the effective Lagrangian
approach.  Furthermore, the manner in which the theory transforms
under both large gauge transformations and under the variation of the
$\theta$ parameter, when monopoles (colorful object) acquire a color
electric charge, are known independently.  Consequently, understanding
the $\theta$ dependence gives us an understanding of confinement, as
$\theta$ acts as a link between colorless and colorful degrees of
freedom.

This paper is an extended version of the letter \cite{JZ99} and is
organized as follows: In Section 2 we overview the properties of the
effective Lagrangian of QCD\cite{HZ}.  Section 3 is devoted to
derivation of the Coulomb gas representation of this effective
Lagrangian, this representation can be interpreted as a dual form of
the original low-energy effective field theory; the massless particles
of the new statistical mechanical problem are identified with
monopoles; the measure for this statistical ensemble is further
analyzed and it is then argued that our particles can be identified
with instanton-quarks. In Section 4 several consequences of the
Coulomb gas representation are derived, in particular, the expression
for the monopole creation-operator (magnetization) is obtained and its
VEV calculated; The magnetic potential in the presence of the domain
walls is also discussed and it is argued that the QCD string can end
on the domain wall.

\section{Effective Lagrangian  and $\theta$ dependence in QCD}

Our analysis begins with the effective low energy $QCD$ action derived
in \cite{HZ}, which allows the $\theta$-dependence of the ground state
to be analyzed and is crucial to the identification of monopole
charges. Within this approach, the Goldstone fields are described by
the unitary matrix $U_{ij}$, which correspond to the $\gmf$ phases of
the chiral condensate: $ \la \overline{\Psi}_{L}^{i} \Psi_{R}^{j} \ra
= - | \la \overline{\Psi}_{L} \Psi_{R} \ra | \, U_{ij}$ with  
\beq
\label{1}
U = \exp \left[ i \sqrt{2} \, \frac{\pi^{a} \lambda^{a} }{f_{\pi}}  + 
i \frac{ 2}{ \sqrt{N_{f}} } \frac{ \eta'}{ f_{\eta'}}  \right] 
 , \quad
U U^{+} = 1 ,
\eeq
where $ \lambda^a $ are the Gell-Mann matrices of $ SU(N_f) $, $ \pi^a
$ is the pseudoscalar octet, and $ f_{\pi} = 133 \; MeV $.  In terms
of $U$ the low-energy effective potential is given by\cite{HZ}:
\beq
W_{QCD}(\theta,U) = - \lim_{V \rightarrow
\infty}~ \frac{1}{V} ~\log
\sum_{l=0}^{p-1} ~\exp \left\{V E \cos \left(- \frac{q}{p} \theta 
+ i\frac{q}{p} \log
{\rm Det}~ U + \frac{2 \pi}{p}~l \right) 
+\frac{1}{2} V \, {\rm Tr} ~( m U +
m^{+} U^{+} ) \right\}.\label{2}
\eeq
All dimensional parameters in this potential are expressed in terms of
the $QCD$ vacuum condensates, and are well known numerically: $ m = {\rm
diag} (m_{q}^{i} | \la \overline{\Psi}^{i} \Psi^{i} \ra | )$; and the
constant $E$ is related to the $QCD$ gluon condensate $ E = \la b
\alpha_s /(32 \pi) G^2 \ra $. The only unknown parameters in this 
construction are the integers $p$ and $q$ which play the same role as
the discrete integer numbers classifying the vacuum states in SUSY
theories. These numbers are related to a discrete symmetry which is a
remnant of the anomaly, and can be found only by explicit dynamical
calculations.  Various arguments can be put forward which support
different values of $p$ and $q$, see e.g. \cite{HZ2,Shifman1};
however, in what follows $p$ and $q$ will not be fixed, rather only
the ratio will be constrained so that in large $N_c$ limit $q/p\sim 1/N_c$
such that the $U(1)$ problem is resolved.

It is possible to argue that equation (\ref{2}) represents the
anomalous effective Lagrangian realizing broken conformal and chiral
symmetries of QCD. The arguments are that:\\
\noindent Equation (\ref{2}):
\def\theenumi{\roman{enumi}}
\def\labelenumi{(\theenumi)}
\begin{enumerate}
\item 
correctly reproduces the VVW effective chiral Lagrangian \cite{Wit2}
in the large $ N_c $ limit 
\\~
[ For small
values of $ ( \theta - i \log {\rm Det} \, U ) < \pi/ q $, the term with $ l
= 0 $ dominates the infinite volume limit.  Expanding the cosine (this
corresponds to the expansion in $ q/p \sim 1/N_c $), we recover
exactly the VVW effective potential \cite{Wit2} together with the
constant term $ - E = - \la b \alpha_s /(32 \pi) G^2 \ra $ required by
the conformal anomaly: \beq
\label{3}
W_{VVW}( \theta, U, U^{+}) = 
-E - \frac{ \la \nu^2 \ra_{YM} }{2} 
( \theta - i \log {\rm Det} \, U )^2
- \frac{1}{2} Tr \, (mU + m^{+}U^{+} ) + \ldots \; . ~~~~~~~
\eeq
here we used the fact that at large $N_c$, $E(q/p)^2= -\la \nu^2
\ra_{YM} $ is the topological susceptibility in pure YM theory.
Corrections in $ 1/N_c $ stemming from Eq.(\ref{2}) constitute a new
result of Ref.\cite{HZ}.]

\item 
reproduces the anomalous conformal and chiral Ward identities of
QCD
\\~[ Let us check that the anomalous WI's in QCD are reproduced
from Eq.(\ref{2}).  The anomalous chiral WI's are automatically
satisfied with the substitution $\theta\rightarrow ( \theta - i \log
{\rm Det} \, U )$ for any $ N_c $, in accord with \cite{Wit2}.
Furthermore, it can be seen that the anomalous conformal WI's of
\cite{NSVZ} for zero momentum correlation functions of the operator $ G^2 $
in the chiral limit $ m_q \rightarrow 0 $ are also satisfied when $E$
is chosen as above.  As another important example of WI's, the
topological susceptibility in QCD near the chiral limit will be
calculated from Eq.(\ref{2}). For simplicity, the limit of $ SU(N_f) $
isospin symmetry with $ N_f $ light quarks, $ m_{q} \ll \Lambda_{QCD}
$ will be considered. For the vacuum energy for small $ \theta < \pi/q
$ one obtains\cite{HZ}
\beq
\label{4}
 E_{vac} (\theta) = -E  + m_q \la \bar{ \Psi} \Psi \ra  N_{f}
\cos \left( \frac{\theta}{N_{f}} \right) + O(m_{q}^2)  \; . 
\eeq
Differentiating this expression twice with respect to $ \theta $
reproduces the result of \cite{SVZ}:
\beq
\label{5}
\lim_{ q \rightarrow 0} \; 
i \int dx \, e^{iqx} \lo T \left\{ \frac{\alpha_s}{8 \pi} 
G \tilde{G} (x)  \, 
\frac{\alpha_s}{8 \pi} G \tilde{G} (0) \right\}  \ro =  
- \frac{ \partial^{2} E_{vac}(\theta)}{ \partial \, \theta^{2}} = 
 \frac{1}{N_f} m_q \la \bar{ \Psi} \Psi \ra  + O(m_{q}^2) \; .
\eeq
Other known anomalous WI's of QCD can be reproduced from Eq.(\ref{2})
in a similar fashion. Consequently, Eq.(\ref{2}) reproduces the
anomalous conformal and chiral Ward identities of QCD and gives the
correct $ \theta $ dependence for small values of $ \theta $, and in
this sense passes the test for it to be the effective anomalous
potential for QCD. ]

\item 
reproduces the known results for the $\theta$ dependence at small
$\theta$\cite{Wit2}, but may lead to a different behavior for large
values $\theta >\pi/q$ if $q\neq 1$ 
\\~
[ As mentioned earlier, at small $\theta \ll\pi/q$ our results are
identical to those found in \cite{Wit2}; the main difference with
\cite{Wit2} arises when $ \theta - i \log {\rm Det} \, U \sim 1$ where cusp
singularities occur.  These singularities are analogous to the ones
arising in SUSY models and show the non-analyticity of the $ \theta $
dependence at certain values of $ \theta $.  The origin of this
non-analyticity is clear, it appears when the topological charge
quantization is imposed explicitly at the effective Lagrangian
level. Thus, the cusp structure of the effective potential seems to be
an unavoidable consequence of the topological charge quantization
(which was not explicitly imposed in the approach of \cite{Wit2}). ]
\end{enumerate}

An interesting note is that in general the $\theta$ dependence appears
in the combination $\theta/N_c$ which naively does not provide the
desired $2\pi$ periodicity for the physical observables; however, such
a behaviour is derived from Eq.(\ref{2}) which does have $2\pi$
periodicity.  This seeming contradiction is resolved by noting that in
the thermodynamic limit, $V \rightarrow \infty$, only the term of
lowest energy in the summation over $l$ is retained for a particular
value of $\theta$, creating the illusion of $\theta/N_c$ periodicity
in observables.  Of course, the values $\theta$ and $ \theta + 2 \pi $
are physically equivalent for the entire set of states, but not for a
selected individual vacuum state.  Consequently, the $\theta$
dependence in the infinite volume limit appearing in the combination
$\theta/N_c$ is a result of being stuck in a particular state.  The
reader is referred to the original papers \cite{HZ} for more detailed
discussions of the properties of the effective potential
(\ref{2}). One final point to be made before moving on is that in
general $E_{vac}(\theta)$ is a multi-valued function with cusp
singularities at $\theta=\frac{(2k+1)\pi}{q}$ where one solution
switches to another.  Furthermore, the number of additional
meta-stable vacuum states is very sensitive to the integer parameters
$p$ and $q$ introduced earlier.  In particular, for the physical
values of the quark masses, there are $q-1$ additional local minima of
the effective chiral potential, which are separated by large barriers
of strength $\sim E$ from the true physical vacuum of lowest energy,
but which are almost degenerate in energy, $\Delta E\sim m$.  Such
availability of almost degenerate vacua imply the appearance of QCD
domain walls and was studied in \cite{FHZ}. The appearance of these
walls have interesting consequences in this work as well.

This section is closed with a few technical but important remarks: The
effective low energy Lagrangian described above has a very special SG
structure.  The fact that such a structure appears in the second term
of Eq. (\ref{2}), $m(U+U^+)\sim \sum_i m_i\cos(\phi_i)$, is quite
natural and is associated with the Goldstone origin of the $\phi_i$
fields. The first term has a similar SG structure $\sim \cos
\left(\left(\sum_i\phi_i-\theta\right)/N_c\right)$; however, its
origin is less trivial and requires some explanation.  Firstly, WI's
imply that the singlet combination $\sum_i \phi_i$ always appears with
$\theta$ in the form $( \theta - i \log {\rm Det}
\, U)$ \cite{Wit2}. Secondly, the appearance of the cosine
interaction, $\cos(\theta/N_c)$, leads to the following
scenario in pure gluodynamics ($\phi_i$'s are frozen): the $(2k)^{\rm
th}$ derivative of the vacuum energy with respect to $\theta$, as
$\theta\rightarrow 0$, is expressed solely in terms of one parameter,
$\frac{1}{N_c}$, for arbitrary $k$:
\beq
 \left.\frac{ \partial^{2k} E_{vac}(\theta)}{ \partial \,
\theta^{2k}}\right|_{\theta=0}
\sim \int \prod_{i=1}^{2k} d^4x_i \left\la  
Q(x_1)...Q(x_{2k})\right\ra \sim \left(\frac{i}{N_c}\right)^{2k},
\label{6}
\eeq
where, $ Q\sim G_{\mu\nu} {\widetilde G}_{\mu\nu}$. This property was
seen as a consequence of Veneziano's solution of the $U(1)$ problem
\cite{Venez}. The reason that only one factor appears in
Veneziano's calculation is that the corresponding correlation
function, $\sim \int \prod_{i=1}^{2k} d^4x_i
\la Q(x_1)...Q(x_{2k})\ra $, becomes saturated at large distances by
the Veneziano ghosts whose contributions factorize exactly, and was
subsequently interpreted as a manifestation of the $\theta/N_c$
dependence in gluodynamics at small $\theta$. However, at that time it
was incorrectly assumed that such a dependence indicates that the
periodicity in $\theta$ is proportional to $N_c$.  Later on \cite{HZ2}
it was argued that the behavior in Eq.(\ref{6}) is a consequence of
the holomorphic structure for the non-perturbative part of the QCD
partition function where the non-perturbative vacuum energy depends
only on a single complex combination
$\tau=1/g^2_0+i\frac{\theta}{32\pi^2}\frac{bq}{p}$ in the same way as
it was in SUSY models\footnote{There is an essential difference
between SUSY models and QCD however: in SUSY models holomorphy is an
exact property of the effective super-potential; however, in QCD it is
only a property of the non-perturbative effective large distance part,
see\cite{HZ2},\cite{FHZ} for more details.}.  The behaviour in
(\ref{6}) was also demonstrated by ``integrating in''a very heavy
fermion and integrating it out afterwards, which must leave the
gluodynamics unaltered \cite{HZ2}. All three independent arguments
support the property (\ref{6}), consequently Eq. (\ref{2}) will be
used as the defining effective action for low energy QCD.

\section{Coulomb Gas Representation of Low Energy Dynamics}

The effective low energy $QCD$ action (\ref{2}) has a very special
quality, all the interactions have a trigonometric form and are in
particular cosinusodal.  This is the defining character of a SG model.
The presence of many fields and many cosine terms with different
harmonics which interact in a highly non-trivial manner serve to
slightly complicate the situation.  Nevertheless, many of the special
properties of the SG theory apply to this model, the admittance of a
Coulomb gas (CG) representation for the partition function is no
different. Although this is a four dimensional theory, and questions
about renormalizability of the theory may come to mind, there are no
such issues here since the effective action is a low energy
one. Following the usual procedure for mapping a statistical CG model
into the field theoretic SG model, the CG picture that arises from the
effective low energy $QCD$ action, Eq. (\ref{2}), will be derived in this
section. The statistical model will be seen to contain several species
of charges which appear due to the presence of several cosine
interactions in the field theory model. The physical meaning of these
charges will be illuminated by applying large gauge transformations on
the partition function and determining the manner in which the charges
transform. This will lead to the identification of one of the charge
species as {\it magnetic monopoles} and of the singlet combination,
$\phi$, in the SG model as {\it the magnetic scalar potential}.  The
remaining charges will also be argued to have magnetic properties; however,
it will prove difficult to construct a precise statement about
them. The statistical model will then be analyzed in subsequent
sections of this paper.

\subsection{Formal Derivation}

Although the mapping between a SG theory and its CG representation is
well known\cite{Po77}, in this section several illustrative steps in
its derivation will be given for completeness. The existence of many
fields and cosine terms only serve to make the formulae more bulky;
however, the basic strategy is the same as in the standard case. Using
the effective potential in Eq. (\ref{2}) the defining partition
function is taken to be,
\beq
\label{Z}
 Z = \Bigg\langle
\exp \Bigg\{- \int d^4 x \Big\{E \cos \left( \frac{q}{p} 
\left(\phi - \theta + 2\pi n \right)  + 2\pi k \right)
+ m_1 \cos (\phi_1)+\dots+m_{N_f}\cos(\phi_{N_f})  
\Big \} \Bigg\} \Bigg\rangle
\eeq
Here, the matrix $U$ characterizing the phases of the chiral
condensate has been placed in diagonal form\footnote{It is well known
that this is the most general form for $U$-matrix describing the
ground state of the system\cite{Wit2}. The off-diagonal elements of
$U$ describe the fluctuations of the physical Goldstones which are
neglected. This is allowed since only the diagonal elements are
relevant in the description of the ground state which is the focus of
this work. This truncation can be justified a posteriori by
demonstrating that the classification of vacuum states based on the CG
representation exactly coincides with the classification based on the
effective Lagrangian approach, see Eq. (\ref{VEV}) where only the
diagonal elements are relevant.  Furthermore, as will be demonstrated
shortly, the most important contribution is related to the singlet
field $\phi = {\rm Tr}~U$ which is unambiguously defined.  All
contributions related to the non-singlet fields are suppressed in the
chiral limit by the quark masses $m_i$.}, $U={\rm
diag}(\phi_1,\dots,\phi_{N_f})$, the combination of fields $\phi =
{\rm Tr}~U$ represents the singlet, $m_a= m_{q}^{a}| \la
\bar{\Psi}^{a} \Psi^{a} \ra |$ denotes the mass of the $a$-th flavor
of quark together with its condensate, $ E = \la b \alpha_s /(32 \pi)
G^2 \ra $ is the vacuum energy of the system, and the angled brackets
represent the Gaussian weighting,
\beq
\langle \dots \rangle = 
\sum_{k=0}^{p-1} \sum_{n=-\infty}^{\infty} \int \D \phi_a ~\dots~
{\rm e}^{ -g^2 \int d^4 x \frac 1 2 ({\vec \nabla} \phi_a)^2}
\label{GaussianWeight}
\eeq 
where $g^2$ is a coupling constant $\sim f_{\pi}^2$  so
that the fields $\phi_a$ are dimensionless and all distances are measured
in the units of $f_{\pi}^{-1}$.

The following strategy for obtaining the CG representation from the SG
model will be employed here: a series expansion in the vacuum energy
and the quark masses will be carried; the cosine terms will then appear
outside of the exponential and are weighted by the Gaussian weighting;
introducing $\Z_2$ valued fields to represent the cosine terms allows
the dynamical fields to re-appear in the exponential, however, now
they appear linearly; in this form it is possible to completely
integrate out the dynamical fields which leads to the final CG
representation. These steps are now carried out in order.  Firstly,
performing a series expansion in $E$ and $m_a$ independently leads to
the following form,
\bea
Z &=&
\sum_{\{M_0,\dots, M_{N_f}\}=0}^\infty \frac{E^{M_0}}{M_0!} 
\frac{m_1^{M_1}}{M_1!}
\dots \frac{m_{N_f}^{M_{N_f}}}{M_{N_f}!} 
\Bigg\langle
\left( \int dx^{(0)} \cos \left( \frac q p 
\left(~\phi(x^{(0)})-\theta+2\pi n\right)+ 2\pi k\right)
\right)^{M_0} \nn \\
&&\hspace{82mm}\times~
\prod_{a=1}^{N_f}\left(
\int dx^{(a)} \cos(\phi_a(x^{(a)}))\right)^{M_a}
\Bigg\rangle \label{seriesexp}
\eea
By introducing the $\Z_2$ valued fields $Q^{(a)}$, for
$a=0,\dots,N_f$, to replace the cosine interactions the dynamical
fields can be integrated out exactly. For $a \ne 0$ the fields are
introduced via, $$\cos(\phi_a)=
\frac{1}{2}\sum_{Q^{(a)}=\pm1}\e^{iQ_a\phi_a} \qquad, \quad a\ne 0.$$
and the charge species $Q^{(a\ne0)}$ will be said to be dual to the
field $\phi_a$. The singlet case, $a=0$, contains additional
parameters,
\beq
\label{cos}
\cos\left(\frac{q}{p}(\phi -\theta+2\pi n)+2\pi k\right) =
\frac{1}{2}\sum_{Q^{(0)}=\pm \frac{q}{p}}
\e^{iQ^{(0)} ((\phi -\theta+2\pi n)+2\pi \frac{p}{q}k) }.
\eeq
and the charge species $Q^{(0)}$ will be said to be dual to the
singlet combination, $\phi= {\rm Tr}~U$. The explicit presence of the
$\theta$-angle in (\ref{cos}) is essential in identifying the physical
meaning of this charge species. Inserting both expressions into the
series expansion of the partition function, Eq. (\ref{seriesexp}),
leads to a form in which the dynamical fields appear at most quadratic
in the exponential and they can be completely integrated out,
\bea
Z &=& 
\sum_{\{M_0,\dots, M_{N_f}\}=0}^\infty \frac{(E/2)^{M_0}}{M_0!} 
\frac{(m_1/2)^{M_1}}{M_1!}
\dots \frac{(m_{N_f}/2)^{M_{N_f}}}{M_{N_f}!}
\int \left(dx^{(0)}_1 \dots dx^{(0)}_{M_0}\right) 
\dots \left( dx^{N_f}_1 \dots dx^{N_f}_{M_{N_f}} \right)
\nn\\&&\hspace{30mm}\times
\sum_{\stackrel{\scriptstyle Q^{(0)}_i = \pm \frac{q}{p}}
{\{Q^{(1)}_i,\dots,Q^{({N_f})}_i\}=\pm 1}}
\Bigg\langle {\rm e}^{i\sum_{b=1}^{M_0} Q^{(0)}_b\left(
\phi(x^{(0)}_b)-\theta+2\pi n + 2\pi \frac{p}{q} k
\right)} 
\prod_{a=1}^{{N_f}} {\rm e}^{i\sum_{b=1}^{M_a} Q^{(a)}_b\phi_a(x^{(a)}_b)} 
\Bigg\rangle 
\label{CGexp}
\eea
The functional integral is trivial to perform and one arrives at the
dual CG action. The brackets can then be removed, as long as the
summations over $n$ and $k$ are restored. After some rewriting the CG
action is,
\bea
S_{CG} &=& i \theta Q^{(0)}_T
- \ln \left(p \sum_{n=-\infty}^{\infty}
{\rm e}^{2\pi i Q^{(0)}_T n} \right)  
+ \frac{1}{2g^2} \sum_{a=1}^{N_f} \Bigg\{
\sum_{b,c=1}^{M_0} Q^{(0)}_b~G(x^{(0)}_b-x^{(0)}_c)~Q^{(0)}_c 
+ 2 \sum_{b=1}^{M_0} \sum_{c=1}^{M_a} Q^{(0)}_b~G(x^{(0)}_b-x^{(a)}_c)
~Q^{(a)}_c \nn\\
&& \hspace{65mm}
 + \sum_{b,c=1}^{M_a} Q^{(a)}_b~G(x^{(a)}_b-x^{(a)}_c)~Q^{(a)}_c 
\Bigg\}. \label{CGaction}
\eea
Here, $Q^{(0)}_T=\sum_{b=1}^{M_0} Q^{(0)}_b$ is the total $Q^{(0)}$
charge for that configuration, $G(x-y)$ denotes the relevant Greens
function of the Laplace operator ($-\Box G(x-y) = \delta(x-y)$), and
the factor of $p$ appears due to the redundancy introduced by the
summation over $k$ in Eq.(\ref{GaussianWeight}).

Notice that the fugacities of the charge species $Q^{(a\ne0)}$ are
given by the masses of the $a^{\rm th}$ quark, while the fugacity of
the charge species $Q^{(0)}$ is proportional to the gluon condensate
$E$.  An important point is that the fugacity of charge species
$Q^{(a\ne0)}$ vanishes in the chiral limit, while that of the
$Q^{(0)}$ charges remains non-zero. A second point is that, in the
chiral limit, this representation does not obviously have invariance
in $\theta$, while in $QCD$ the $\theta$-angle appears with quark
masses and hence disappears in this limit. It is possible to
demonstrate that the partition function has this property; however,
because its proof is rather technical, the details are deferred to
appendix A.

There are several important features of the action (\ref{CGaction})
which should be noted. Firstly, the summation over $n$ forces the
total $Q^{(0)}$ charge, $Q_T^{(0)}$, to be an integer. Such a
constraint is the analog of the quantization of the topological charge
and is to be expected. Additionally, since this species has charges
$\sim\frac{q}{p}\sim 1/N_c$, this constraint enforces a fractional
quantization on the total charges: the difference in the number of
positive and negative charges of species $Q^{(0)}$ must be an integer
multiple of $N_c$. Secondly, the $\theta$-angle only interacts with
the $Q^{(0)}$ species. This directly results from the $Q^{(0)}$
species being dual to the singlet field $\phi$, which is the only
field directly interacting with the $\theta$-angle. Finally, the
$\theta$-dependence acts only to supply an overall phase factor for
each configuration and leads to the very natural interpretation of
non-trivial $\theta$-angles as introducing an overall background
charge. Turning attention to the interactions amongst the charges, the
species $Q^{(0)}$ is seen to interact with all species $Q^{(a)}$
($a=0,\dots,N_f$); however, the other species, $Q^{(a\ne0)}$, only
interact with their own species and species $Q^{(0)}$. This peculiar
behave is, once again, due to species $Q^{(0)}$'s association with the
singlet while the other charge species are associated with particular
components of the chiral condensate. It is interesting that the
structure of interactions between the charges resembles the 't~Hooft
``determinant'' interaction, in which all $N_f$ fermions form a vortex
with $N_f$ legs, while different species do not notice one
another. Such a structure can be easily understood in the instanton
picture, in which the existence of exactly one zero mode for each
given flavor provides precisely this kind of interaction.  In the
present context, however, a straightforward explanation in terms of
the original gauge and matter degrees of freedom is not
available. Nevertheless, the interactions occurring in
(\ref{CGaction}) are quite similar to other known statistical systems,
yet they do not literally correspond to interactions appearing in any
previously discussed model.

\begin{figure}[t]
 \epsfysize=4cm
\hspace*{10mm}\epsffile{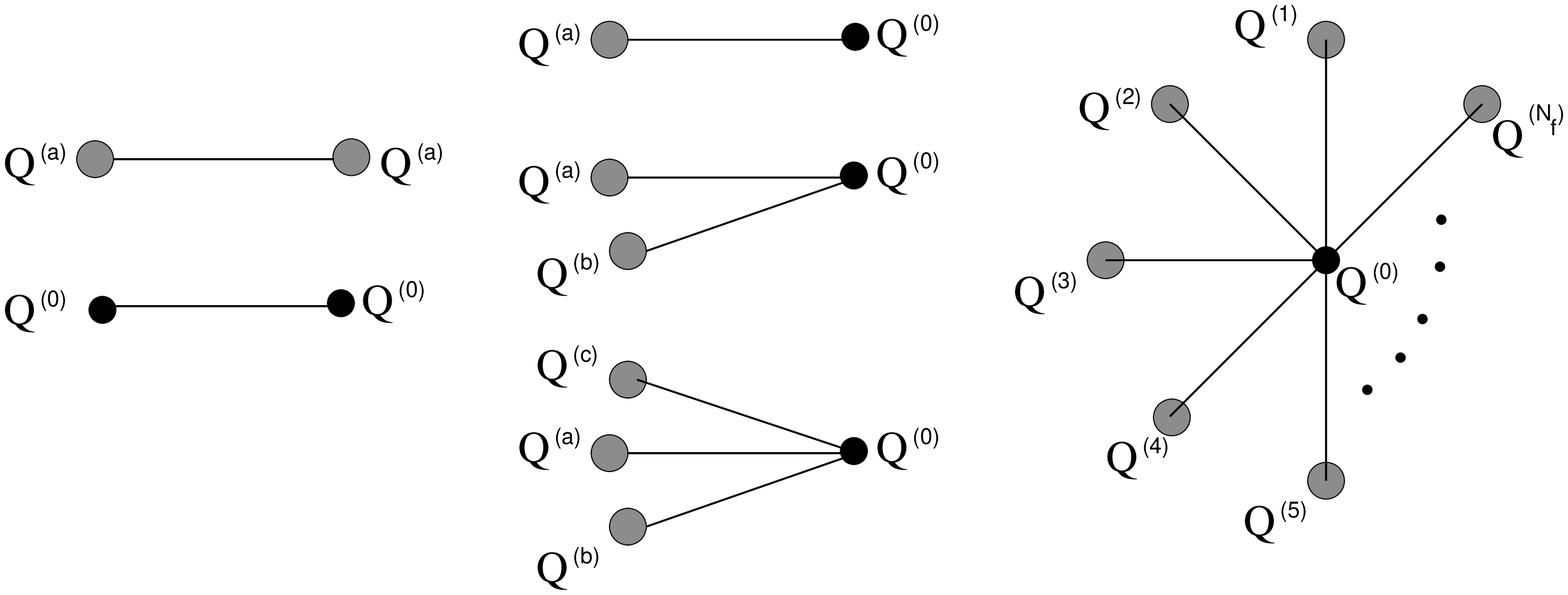}
\caption{The interactions amongst the charges in the CG representation. 
In this diagram $a\ne b\ne c\ne0$. The charges can interact only
amongst themselves (diagrams furthest left), or with the $Q^{(0)}$
charge.
\label{interactions}}
\end{figure}

Figure \ref{interactions} summarizes the manner in which interactions
take place in our statistical system. Notice that the $SU(N_f)$
symmetry is restored in the limit of equal quark masses. This occurs
since the fugacities of all charge species are identical in this
limit, and the multiplicity of any of the diagrams depends only on the
number of legs and not on what charges are attached to them.

\subsection{Physical Interpretation of Charges}

Expression (\ref{CGaction}) clearly shows that the statistical
ensemble of particles interact according to the Coulomb law $\sim
|x-y|^{-1}$ in static (time-independent) configurations ( $\sim
|x-y|^{-2}$ for time varying ones).  An immediate suspicion following
from this observation is that these particles carry a magnetic and/or
electric charge, since charges of that type interact precisely in the
above manner. Due to the direct interaction of the $\theta$-angle and
the charges, $Q^{(0)}$, associated with the singlet field, $\phi$, it
is quite plausible that they may in fact have a magnetic nature.  This
suspicion will be corroborated in a moment. Furthermore, since there
are several species of charges in the CG picture, a natural question
is, ``What is the physical relevance of the remaining flavor
non-singlet charges $Q^{(a\ne0)}$?'' Evidence of their magnetic nature
is found by noting that the VEV's, $\la \phi_a(\theta)\ra$, of these
fields do in fact depend on the $\theta$-angle. The implication is
that {\it all} charge species, not only the one dual to the singlet
field, have a magnetic origin and are connected to
monopoles. Unfortunately, a precise relation of the charges dual to
the non-singlet fields and monopole charges is still missing.

As alluded to above, a precise relationship of the $Q^{(0)}$ charges
and magnetic monopoles can be made. However, in order to prevent a
long interruption in the presentation, only the simplest
identifications will be given here, while the interested reader is
referred to Appendix B for further detailed discussions. The charge
$Q^{(0)}$ was originally introduced in a very formal manner so that
the QCD effective low energy Lagrangian (\ref{2}) can be written in
the dual CG form (\ref{CGaction}). Now the physical content behind
these manipulations will be given.

To begin, consider the $SU(2)$ Georgi-Glashow model in the weak
coupling regime, with a $\theta$-term, and let the scalar $\Phi^a$
have a large VEV.  The monopole solution can be constructed explicitly
and the so-called Witten effect \cite{Witten1}, where the monopole
acquires an electrical charge, takes place\footnote{It should be
noted that the following arguments hold not only for the
Georgi-Glashow model. Monopoles could appear in the system via a
nontrivial dynamical effect and are not necessarily described in the
terms of the original fields in the weak coupling regime. For example,
they can appear as a result of gauge fixing \cite{Hooft1} which is
certainly not the unique gauge-invariant construction. Nevertheless,
it is believed that the Witten effect takes place for such 't Hooft
monopoles \cite{Hooft1}, as well as for any other monopoles. The most
important feature of the Witten effect are the properties of large
gauge transformations rather than specific features of the theory and
matter contents.}. Let $N$ denote the generator of large gauge
transformations corresponding to rotations in the $U(1)$ subgroup of
$SU(2)$ picked out by the gauge field, i.e. rotations in $SU(2)$ about
the axis $n^a=\frac{ \Phi^a}{|\Phi^a|}$.  Rotations by an angle of
$2\pi$ about this axis must yield the identity for arbitrary
configurations, which implies \cite{Witten1} that the magnetic
monopoles carry an electric charge proportional to $\theta$. Indeed,
\bea
\label{7}
1= e^{i2\pi N}=e^{i2\pi\frac{Q}{e}- i\theta \frac{e M}{4\pi}},
\eea
where, 
\bea
\label{8}
  M=\frac{1}{v}\int d^3xD_i\Phi^aB_i^a ,\quad
   Q=\frac{1}{v}\int d^3xD_i\Phi^aE_i^a, 
\eea
are the magnetic and electric charge operators respectively, expressed
in terms of the original fields, and $v$ is the VEV of $\Phi^a$ at
infinity. The combination $\frac{eM}{4\pi}=n_m$ in Eq. (\ref{7}) is an
integer and determines the magnetic charge of the configuration.  As
usual, it is assumed that (\ref{7}) remains correct in the strong
coupling regime when $v$ is not large and/or in the more radical case
when $\Phi^a$ is not present in the original formulation. Indeed, as
explained in \cite{Hooft1} the existence of $\Phi^a$ is not essential
and some effective fields may play its role.  One finds that monopoles
do exist and the Witten effect expressed by formula (\ref{7}) remains
unaltered even when monopoles appear as singularities in the course of
the gauge fixing procedure as described in \cite{Hooft1}.

Restricting attention to terms which are proportional to the
$\theta$-parameter, a comparison between the CG representation,
Eq.(\ref{CGaction}), and Eq.(\ref{7}) will now be carried out. From
the CG the relevant term is the total charge, $Q^{(0)}_T$, of the
configuration, while in Eq.(\ref{7}) the relevant factor is the total
magnetic charge $\frac{e M}{4\pi}$ for each time slice. The following
identification is then made,
\bea
\label{9}
Q^{0}_T=\frac{e M}{4\pi}= n_m \in {\rm Z\!\! Z}.
\eea 
A non-trivial check on this identification can be made by noticing
that the $Q^{(0)}$ species are fractionally charged $Q^{(0)}\sim
1/N_c$, while the total charge $Q^{(0)}_T $is forced to be integer.
This follows from the constraint that the difference in the number of
positive and negative charges of species $Q^{(0)}$ must be an integer
multiple of $p\sim N_c$ (see discussions after (\ref{CGaction})) and
this behaviour is identical to the that of the total magnetic charge.

From these simple observations one can immediately deduce that our
fractional magnetic charges $Q^{(0)}$ cannot be related to any
semi-classical solutions, which can carry only integer charges;
rather, configurations with fractional magnetic charges should have
pure quantum origin.  Of course, this is a simplified explanation of
the identification between the charges $Q^{(0)}$ from (\ref{CGaction})
and the physical magnetic charges. To make this correspondence more
precise the transformation properties of each relevant degree of
freedom under a large gauge transformation must be computed and the
details of such a calculation can be found in Appendix B. 

As already mentioned, the charges $Q^{(a\ne0)}$ are also suspected to
carry magnetic charge, however, a precise statement cannot be made
because the large gauge transformations are sensitive only to the
singlet combination of fields $\phi$ and consequently only to the dual
charge $Q^{(0)}$.

It is interesting to note that a similar phenomenon in the
two-dimensional $O(3)~\sigma$-model (more generally, in $CP^{N}$
models) has been known for a long time\cite{Fateev}. Namely, an exact
accounting and resummation of the $n$-instanton solutions maps the
original problem to a $2d$-CG with fractional charges (the so-called
instanton-quarks); however, the total topological charge of each
configuration is always integer. In that case, it is clear that the
quantum fluctuations completely reconstruct all degrees of freedom:
each instanton is essentially a superposition of $N_c$
instanton-quarks such that, locally, they appear as fractionally
charged particles.  This superposition is quite nontrivial; however,
it is stable under small fluctuations and eventually becomes the only
relevant degree of freedom. Indeed, the corresponding two-dimensional
CG of the instanton-quarks can be rewritten as a quantum field theory
with effective Lagrangian given by the SG model where the dynamics is
carried by some effective field $\phi$ \cite{Fateev}.  This effective
SG Lagrangian describes confinement (mass-gap) and many other
important properties of the $O(3)~\sigma$-model quite well. Our CG
system with fractional magnetic charges, $Q^{(0)}\sim 1/p\sim 1/N_c$,
is very similar to the situation described above. The difference with
\cite{Fateev} is the starting point. In the case of the
$O(3)~\sigma$-model the statistical-mechanical problem was reduced to
some low-energy Lagrangian. However, in the present case we start from
the low-energy Lagrangian and a statistical-mechanical model of
fractional magnetic monopoles is derived.

Until now, the integers $p$ and $q$ have been considered as free
parameters in the theory, and can only be fixed by an explicit
dynamical calculation. The only constraint on these parameters has
been that in the large $N_c$ limit their ratio satisfies $q/p \sim
1/N_c$ so that the $U(1)$ problem is resolved. An argument which
supports a particular choice for these arguments will now be
given. Recall that $q/p$ is related to the $\theta$ dependence of
physical observables (such as the vacuum energy $ E_{vac}(\theta)$ as
can be seen from Eq.(\ref{6})).  In particular, an observable with
expectation value $\sim \exp(i\frac{\theta}{N_c})$ corresponds to
$q=1$, $p=N_c$.  The analysis of softly broken SUSY theories
\cite{Shifman1} strongly suggests that this choice of parameters, 
$q=1$ and $p=N_c$, is in fact correct even for non-supersymmetric
models. In this case the number of integrations over $d^4x_i^{(0)}$ in
Eq.(\ref{CGexp}) exactly equals $4 N_c k$, where $k$ is an arbitrary
integer (not to be confused with the summation parameter appearing in
the partition function). This is a direct consequence of the fact that
the number of species $Q^{(0)}$ must be an integer multiple of
$N_c$\footnote{To be more precise, the difference in the number of
positive and negative charges of species $Q^{(0)}$ must be an integer
multiple of $N_c$. However, in what follows charges with a specific
sign, say positive, will be identified with instanton-quarks.  Such an
identification will be made by only retaining a single term in the
expansion of the cosine function in (\ref{cos}) to obtain a
statistical model with only like signed charges analogous to
(\ref{CGexp}, \ref{CGaction}). In this case the number of particles,
and not the difference in the number of positive and negative charges,
must be an integer multiple of $N_c$. Restricting the model to contain
only like charged particles is slightly illegal for $\theta=0$ in the
presence of light quarks, because in the large volume limit the
corresponding contribution to the partition function is suppressed by
the inverse volume of the system (due to the neutrality requirement)
and/or by the quark mass.  However, the question regarding a measure
of the corresponding configuration (before integration over
$d^4x_i^{(0)}$) is a perfectly acceptable one which has an answer. In
general, it is expected that both instantons and anti-instantons must
be present in the system in order to give a nonzero result for the
partition function in the chiral limit (see discussions at the end of
this section). In this respect there is a difference with the
$O(3)~\sigma$-model without fermions where exclusively instantons
could provide a nonzero contribution to the partition function.}.
This number, $4 N_c k$ exactly corresponds to the number of zero modes
in the $k$-instanton background\cite{Instanton}, and, correspondingly,
to a number of collective variables associated with these zero modes.
In other words, $4 N_c k$ coordinates $x_i^{(0)}$ describe the
collective coordinate integration measure of the $k$-instanton
solution.

Motivated by the analysis of the two-dimensional $\sigma$-model in
\cite{Fateev} and by the above observation regarding the number of
collective variables $4 N_c k$ mentioned above, we {\it conjecture}
that at large distances our particles, $Q^{(0)}$, with fractional
magnetic (and electric for $\theta\neq 0$) charges are indeed the
instanton-quarks (i.e. these are related to instantons) suspected long
ago \cite{Belavin}. One immediate objection to this conjecture is that
since it has long been known (see e.g. \cite{Shuryak}) that instantons
can explain most low energy QCD phenomenology (chiral symmetry
breaking, resolution of the $U(1)$ problem, spectrum, etc) with the
exception of its most important property --confinement; and we claim
that our magnetic monopoles, $Q^{(0)}$, are instanton-quarks and yet
also claim that confinement arises in this picture (they will
Bose-condense, see section IV); how can this be consistent?  This
seeming objection is answered by noting that it in the dilute gas
approximation, when the instantons and anti-instantons are well
separated and maintain their individual properties (sizes, positions,
orientations), quark confinement can not be described.  However, the
lessons from the two-dimensional $ \sigma $ model \cite{Fateev}
teaches us that in strongly coupled theories the instantons and
anti-instantons lose their individual properties (instantons will
``melt'') their sizes become very large and they overlap. If this
happens, the description in terms of the instantons and
anti-instantons is not appropriate any more, and alternative degrees
of freedom should be used to describe the physics. The relevant
description is that of instanton-quarks (which, according to our
conjecture, are particles with monopole charges $Q^{(0)}$). Further to
this point, lattice simulations do not contradict this picture where
large instantons induce the magnetic monopole loops forming large
clusters, see e.g.\cite{Diakonov} and references therein. Also, the
connection between monopoles and instantons on the classical level is
not a very new idea \cite{nahm}.  Indeed, quite recently, such a
relation was established for the periodic instantons (also called
calorons) defined on $R^3\times S^1$ in the presence of a Wilson line
\cite{vanbaal}.  Furthermore, a similar relation was seen in the study
of Abelian projection for instantons \cite{Polikarpov,Brower}, albeit
at the classical level.  In particular, Brower et. al. \cite{Brower}
demonstrated that the instanton's topological charge, $Q$, is given in
terms of the charge, $M$, of the monopole, which forms the monopole
loop, by the expression $Q=\frac{eM}{4\pi}$.  This formula is very
similar to our relation (\ref{9}), where the total topological charge,
$Q^{(0)}_T$, for a configuration containing a number of particles,
$\{Q^{(0)}_i\}$, described by the system (\ref{CGaction}) was
identified with the total magnetic charge for each time slice for the
same configuration.  In spite of this similarity, the physical
interpretation of this relation is quite different for these two
cases. In the former case it is an identity for a configuration with
integer monopole and topological charges satisfying the classical
equations of motion; while in the latter case it is a relation for a
configuration of the fractionally charged monopoles and topological
charges (fractional) for the instanton quarks (which presumably appear
after integration over quantum fluctuations in the multi-instanton
background).  However, given that the instantons at the classical
level, and the instanton-quarks at the quantum level both have similar
relations suggests that our conjecture is possibly correct. Additional
supporting arguments based on the instanton measure for arbitrary
gauge groups are given in our concluding remarks.

As a last remark regarding our conjecture, note that the QCD effective
Lagrangian (\ref{Z}), at $\theta=0$, is invariant under $CP$
transformations; consequently, the CG representation (\ref{CGexp}) is
also invariant under $CP$ at $\theta=0$.  This implies that only when
instantons and anti-instantons are taken into account can the
statistical ensemble of charges be equivalent, at large distances, to
a statistical ensemble of instanton-quarks. If either are left out of
the description $CP$ symmetry is broken.  The same conclusion also
follows from the fact that the partition function, (\ref{CGexp}), is
non-zero in the chiral limit\footnote{However, only one species of
particles with charges $Q^{(0)}$ contribute to the partition function
in this case; other species with charges $Q^{(a\ne0)}$ have
zero fugacities in the chiral limit and, therefore, are not
present in the system.}  $m_a=0$ . Such a behaviour of the partition
function implies that our system (\ref{CGexp}) can be equivalent to
the statistical ensemble of the instanton-quarks if and only if both
instantons and anti-instantons are present in the system, and there is
no excess of the topological charge (for $\theta=0$) which, otherwise,
would lead to the chiral suppression of the partition function.

\section{Operators}

In the previous section, a CG representation which describes the low
energy dynamics of the $QCD$ action was derived. The charges were
found to have magnetic properties, and the charge species, $Q^{(0)}$,
dual to the singlet field, $\phi$, was explicitly identified as
magnetic monopoles. However, the formal mapping from the SG
description  to the CG description, (\ref{2}) to
(\ref{CGaction}), is not very useful if the dynamics of the fields and
charges are not understood.  In order to understand the long-distance
features of the theory, the manner in which the various charge
species behave must be investigated. In particular, if the magnetic
charges Bose-condense, this indicates the onset of quark
confinement. To investigate the possibility for such a condensation
an expression for the magnetic charge creation operator, $\MM$, must
be found and its VEV (magnetization) calculated. In this section the
magnetization $\la \MM\ra $ will be demonstrated to be non-vanishing;
consequently, the theory is in the confining phase.  Of course, this
is not a proof of confinement in $QCD$ because the effective action
was constructed under the assumption that the system is in the
confining phase, i.e. (\ref{2}) contains only colorless degrees of
freedom.  In spite of this, the result is quite nontrivial and can be
considered as a self-consistency check of our identifications. In
addition, it provides a very nice intuitive picture of the origin of
the confinement in $QCD$.

\subsection{Charge Creation}

Consider inserting a charge, $\q_0=\pm\frac{q}{p}$, of species
$Q^{(0)}$ in the bulk at the point $\X$.  To accommodate this
insertion, the partition function must be altered by restricting, say,
the coordinate $x^{(0)}_0=\X$ and $Q^{(0)}_0 = \q_0$. Tracing the
steps in the previous section in reverse order with this restriction
leads to the appropriate operator equivalence. However, it is simpler
state the result and then demonstrate that it is correct. The operator
equivalence, which is proven correct shortly, is,
\beq
\MM(\q_0,\X) = {\rm e}^{i\q_0\left( 
\phi(\X) - \theta+2\pi n + 2\pi \frac{p}{q} k\right)},
\label{Q0insert}
\eeq 
where $\MM$ denotes the operator which inserts the charge in the bulk.
The singlet combination $\phi$ naturally appears here since $Q^{(0)}$
is dual to $\phi$, while the appearance of $\theta, n$ and $k$ appear
due to the manner in which $Q^{(0)}$ was introduced originally (see
Eq. (\ref{cos})).  Clearly this operator must be inserted under the
summations over $n$ and $k$. On inserting (\ref{Q0insert}) into the
partition function and performing the series expansion in $E$ and
$m_a$ then introducing the $\Z_2$ valued fields, as before, the
modified CG action is arrived at,
\bea
{\rm e}^{-S_{CG}} &=& 
\Bigg\langle \left({\rm e}^{i\sum_{b=1}^{M_0} Q^{(0)}_b\left(
\phi(x^{(0)}_b)-\theta+2\pi n + 2\pi \frac{p}{q} k
\right)}
\prod_{a=1}^{{N_f}} {\rm e}^{i\sum_{b=1}^{M_a} Q^{(a)}_b\phi_a(x^{(a)}_b)} 
\right)
{\rm e}^{i\q_0\left( 
\phi(\X) - \theta+2\pi n + 2\pi \frac{p}{q} k\right)} \Bigg\rangle 
\nn
\eea
The term appearing outside the inner parenthesis is the suggested
operator for the creation of monopoles. Since $\X$ and $\q_0$ are not
being summed over, this term will act as a source for a fixed charge
at a fixed position. Collecting this term with the first term and
performing the functional integrals over the scalar fields,
demonstrates that $\q_0$ will interact within the statistical model
precisely in the form of charge species $Q^{(0)}$ with the exception
that its charge and position is not being summed over. Consequently,
the operator relation (\ref{Q0insert}) is proven to hold.

The insertion of any other charge species $Q^{(a\ne0)}$ can be found
via an analogous method, and the operators for these charge
species are found to be,
\beq
\MM({\q_a},\X) = {\rm e}^{ i \q_a\phi_a(\X)}
\label{Qainsert}
\eeq

Notice that (\ref{Q0insert}) explicitly depends on $\theta$, $n$, $k$
and the singlet combination $\phi$, while (\ref{Qainsert}) only
depends on the scalar field of its own species, $\phi_a$, and not on
any other parameters in the theory. This is of course a consequence of
$Q^{(0)}$ being associated with monopoles and must transform
correspondingly under large gauge transformations. Furthermore, the
manner in which the singlet field, $\phi$, appears in (\ref{Q0insert})
allows the further identification of $\phi$ in the CG representation
as {\it the magnetic scalar potential}.

Now that the relevant operator relations have been identified,
questions about the dynamical nature of the system can be asked. The
most pressing one being, ``Does the operator which creates a charge
have a non-zero expectation value?''  This is a difficult question to
answer in the CG representation, however, in the dual SG representation
an answer at the semi-classical level can be given. Its solution,
however, requires a knowledge of the VEV's of the phases of the chiral
condensates, $\la \phi_a \ra$, such that at the semi-classical level $
\la\MM\ra \simeq {\rm e}^{ i \q_a\la \phi_a\ra}$, and the problem is
reduced to the calculation of $\la \phi_a\ra$.  These VEV's can be
found by the minimization of the effective potential (\ref{Z}), and are
given by the equations\cite{HZ}, \bea
\label{phases}
\sin\left(\frac{q}{p}(\la\phi\ra-\theta)\right) = -\frac{p}{q}\frac{m_a}{E} 
\sin \la\phi_a\ra , ~~~~~~~~\phi\equiv{\rm Tr}~U=\sum_{a=1}^{N_f}\phi_a,
 ~~~~~~~~a=1, \dots N_f
\eea
where the potential has been chosen to be in the particular branch
$k=0$. In the limit of equal quark masses, $m_a=m$, and assuming
$\frac{m}{E} \ll 1$, the system can be solved perturbatively, and to
lowest order the VEV's are: $\la \phi_a\ra \simeq \frac{\theta}{N_f}$
for small $\theta$.  Earlier on it was mentioned that the VEV's of {\it
all} components of the chiral condensate acquire dependence on the
$\theta$-angle and this is now explicitly demonstrated. The suspicion
of the magnetic origin of the non-singlet fields can now be justified
somewhat, although an explicit construction, as in the case of the
singlet field, is still missing. By inserting these VEV's into the
expressions for the charge creation operators the magnetization at the
semi-classical level can be evaluated,
\bea
\la \MM(\q_a, \X) \ra \sim \left\{  
\begin{array}{ll}
1 & a = 0 \cr
e^{i\frac{\theta}{N_f}} & a\ne 0.
\end{array}\right.
\label{VEV}
\eea 
From Eq. (\ref{VEV}) one can clearly see that operator $\MM$ is the
order parameter of the system and its nonzero VEV implies the
condensation of monopoles (dyons for non-zero $\theta$). Eventually
this condensation leads to  confinement (oblique confinement for
non-zero $\theta$) in the system.

To close this section note that the phase of the order parameter $\la
\MM\ra$ (\ref{VEV}) which labels different vacua in the CG representation
(\ref{CGexp}) is determined by the same equation (\ref{phases})
describing the chiral condensate phases in SG representation (\ref{Z}),
which is the conventional description of low-energy QCD in terms of
the chiral condensates.  The physical meaning of the phases, $\phi_a$,
presented in the CG representation (\ref{CGexp}) and in the effective
Lagrangian approach (\ref{Z}) is, however, quite different. In the
former case they represent the magnetic scalar potential; while in the
latter case they are the phases of the chiral condensate. It is quite
remarkable that in spite of the very different physical
interpretations suggested by the two (dual) approaches lead to the
same classification of vacua. One final remark is that
generalizations of our calculations of $\la \MM(\q_a, \X) \ra$ for the
case of different branches, arbitrary $\theta$ and non-equal quark
masses is quite straightforward, and reduces to the solution of
Eq. (\ref{phases}) for the phases $\la \phi_a \ra $.  The
corresponding analysis has been carried out in \cite{HZ} and it is not
repeated here.

\subsection{Background Fields and Wilson Loop Operator}

As demonstrated in the previous section, the VEV of magnetization is
non-zero, $\la \MM({\q_0},\X)\ra \neq 0$; therefore, our system is in
the confinement phase as expected.  Consequently, the VEV of the
Wilson loop $\la W\ra$ must show an area law dependence, and it is
interesting to find an explicit form of the corresponding Wilson loop
insertion in the SG or CG representation.  Naively this seems like a
hopeless goal, because the starting point, the effective low-energy
QCD Lagrangian, does not contain any fundamental degrees of freedom
(gluons and quarks) in terms of which the Wilson operator is defined.
Nevertheless, as will be seen in a moment, the corresponding Wilson
loop insertion operator can be recovered.  The key point in this
correspondence is again the identification of our charges $Q^{(0)}$
with the physical magnetic charges, Eq. (\ref{9}). Once this
identification is made, the singlet field, $\phi$, in the expression
(\ref{Q0insert}) for the magnetic creation operator is interpreted as
the {\it magnetic scalar potential} such that $\MM({q_0},\X) = {\rm
e}^{ i q_0\phi(\X)}$ represents the interaction energy of a monopole
of (fractional) charge $q_0$ at position $\X$ in the presence of the
potential of other monopoles.  With this identification it is quite
obvious what kind of replacement in our formulae (\ref{Z}) and
(\ref{CGexp}) should be made in order to insert a Wilson loop
operator.  Physics suggests the following picture: For each given
time-slice, $x_0$, the Wilson loop inserted (at $(x_1,x_2) $ plane) into
the monopole plasma behaves like a sheet of magnetic dipoles, and
produces at large distance from the sheet a dipole magnetic field.
The monopoles of the vacuum plasma, however, are polarized by this
dipole field and react to produce a dipole field to try to cancel the
field from the sheet.  Therefore, this interaction of between a source
with magnetic charges from the system changes the magnetic scalar
potential $\phi$.  This change, however occurs only in the thin region
where the magnetic plasma of the system does not cancel the field from
the sheet. The thickness of this region is order $E^{-1}$, the only
dimensional relevant parameter in the system.  Therefore, the Wilson
loop insertion changes the magnetic scalar potential, $\phi$, by
$\phi+\eta$ where $\eta$ is an addition to the magnetic scalar
potential due to a dipole layer of unit strength on the sheet $S$
represented by the Wilson surface. The key point is that even though
gauge fields are not present in our description, through the
identifications of the charges as magnetic monopoles, just enough
information about them is encoded in the magnetic scalar potential,
$\phi$, to obtain interesting properties. The transformation property
of $\phi$ under the Wilson loop insertion, as explained above, is
$\phi\rightarrow\phi+\eta$.  The function $\eta(x)$ has a
discontinuity of $4\pi$ when the point $x$ crosses the surface bounded
by the Wilson loop for each given time slice.  In a real physical
situation the dipole layer is not infinitely thin (as mentioned above
it is order of $E^{-1}$); however, incorporating the finite
thickness is beyond the present analysis.

An explicit representation for the appropriate source term will now be
constructed. It will be explicitly demonstrated that such a source
interacts with the magnetic charges $Q^{(0)}$ exactly in the same manner
as the $\phi$ field, $\sim {\rm e}^{ i Q^{(0)}\eta}$. Therefore,
the $\eta$ source insertion can be identified with an appropriate
change in the magnetic scalar potential $\phi\rightarrow\phi+\eta$.
But, as explained above, the Wilson loop insertion must also lead to
precisely this kind of shift $\phi\rightarrow\phi+\eta$ in SG
(\ref{Z}) or CG (\ref{CGexp}) representation.

The insertion of background dependence is furnished by altering the
Gaussian weighting (\ref{GaussianWeight}) to include source terms,
\bea
\langle \dots \rangle = 
\sum_{k=0}^{p-1} \sum_{n=-\infty}^{\infty} \int \D \phi_a ~\dots~
{\rm e}^{ -g^2 \int d^4 x \left\{ 
\frac 1 2 ({\vec \nabla} \phi_a)^2 + \phi_a \Box \eta_a^{(0)}\right\}}
\nn
\eea 
The classical equations of motion are of course modified by this
insertion, and with appropriate choices of the source fields,
$\eta_a^{(0)}$, the classical solution can be made kink like, for
example.  However, $\eta_a^{(0)}$ will not be constrained at this
moment, and any source (Wilson loop, domain wall or anything else) can
be described by this modification.  Of course, it is interesting to
study what this modification implies for the CG representation. On
expanding the partition function in powers of $E$ and $m_a$, and
introducing the $\Z_2$ valued charge fields one  arrives at precisely
equation (\ref{CGexp}) where the angled brackets are the modified ones
above. Performing the functional integral posses no difficulty and one
arrives at the CG representation,
\bea
Z &=& 
\sum_{\{M_0,\dots, M_{N_f}\}=0}^\infty \frac{(E/2)^{M_0}}{M_0!} 
\frac{(m_1/2)^{M_1}}{M_1!}
\dots \frac{(m_{N_f}/2)^{M_{N_f}}}{M_{N_f}!}
\int \left(dx^{(0)}_1 \dots dx^{(0)}_{M_0}\right) 
\dots \left( dx^{(N_f)}_1 \dots dx^{(N_f)}_{M_{N_f}} \right)
\nn\\&&\hspace{10mm}\times
\sum_{\stackrel{\scriptstyle Q^{(0)}_i = \pm \frac{q}{p}}
{\{Q^{(1)}_i,\dots,Q^{({N_f})}_i\}=\pm 1}}
{\rm e}^{-S_{CG}-\frac{g^2}{2}\int d^4 x~\eta_a\Box\eta_a}
{\rm e}^{i\sum_{b=1}^{M_0} Q^{(0)}_b \eta(x^{(0)}_b)}
\prod_{a=1}^{N_f}{\rm e}^{i \sum_{b=1}^{M_a} Q^{(a)}_b \eta_a(x^{(a)}_b)}
\nn
\eea
In the above, $S_{CG}$ is the CG action containing the interactions of
the charges and is unaltered from its form given in equation
(\ref{CGaction}), also $\eta\equiv \eta_1+\dots+\eta_{N_f}$.  This
then allows the identification of the operator, in the CG
representation, which introduces a source for the classical equations
of motion of the phases of the chiral condensates,
\beq
\C(\eta_a) = {\rm e}^{-\frac{g^2}{2}\int d^4 x~\eta_a\Box\eta_a}
~{\rm e}^{i\sum_{b=1}^{M_0} Q^{(0)} \eta(x^{(0)}_b)}
\prod_{a=1}^{N_f} {\rm e}^{i\sum_{b=1}^{M_a} Q^{(a)}_b 
\eta_a(x^{(a)}_b)} \label{bkfield}
\eeq
in this formula the term $ {\rm e}^{-\frac{g^2}{2}\int d^4
x~\eta_a\Box\eta_a}$ represents a self-interaction of the source with
itself, it is irrelevant for the present purposes and will be omitted
in the analysis.  The second term in (\ref{bkfield}) is much more
interesting and gives a very simple prescription for the insertion of
any source. If one wants to insert some classical field configuration
(source) in the original formulation, then in the CG picture one need
only introduce a phase factor which gives rise to a source term for
that background - all interactions among the charges remain unaltered.
The charges simply interact with the source field just as if it were
an external magnetic field, each charges species, $Q^{(a\ne0)}$,
interacts with the source component with which it is associated to,
$\eta_a$, while the singlet charge species $Q^{(0)}$ interacts with
all of the background fields. The special form of the SG action is
what admits this very simple picture and is at the heart of the CG
representation itself.  Consequently, it is expected that the analysis
carried out here has wide applicability and can be applied for any
kind of sources.

In deriving (\ref{bkfield}) no assumptions about the source fields
$\eta_a$ have been made. The remainder of this section will be devoted
to the most interesting source - the Wilson loop operator
insertion. As argued above, any source can be accounted for by the
shift $\phi\rightarrow\phi+\eta$, and the Wilson loop operator
insertion is no exception. Our problem, therefore, is reduced to the
calculation of a difference in magnetic scalar potential due to the
insertion of this specific source. Before dealing with the $4d$ case,
some relevant formulae from the $3d$ case, where the physics is well
understood, will be reviewed; then the appropriate generalizations to
the $4d$ case will be made. The reason that the $3d$ analysis can be
easily generalized to the $4d$ case is that in both cases the
interactions between particles in the CG representation
(\ref{CGaction}) is determined by the relevant Greens function
($1/|x-y|$ or $1/(x-y)^2$ respectively); consequently, the Wilson loop
insertion in the CG representation should be expressible in terms of
the appropriate Greens function.  Hence, if the Wilson loop insertion
operator in $3d$ can be expressed exclusively in terms of the $3d$
Greens function, $1/|x-y|$, the transition from three to four
dimensions will be furnished by the replacement of the $3d$ Greens
function by the $4d$ Greens function such that the static limit is
reproduced.  In addition, it is expected that the Wilson loop
insertion is a Lorentz-covariant expression.

Recall that in the well-understood $QED_3$, Polyakov\cite{Po77}
demonstrated that the expectation value of a Wilson loop operator can
be written in terms of a CG with the insertion of the operator,
\beq
\label{wilson3}
\la W \ra \sim {\rm e}^{ \frac{i}{2}\eta (x, C)},~~~
\eta(x, C)=\int_{\Sigma_C} d^2\sigma_{\mu}(y )\frac{(x-y)_{\mu}}{|x-y|^3}=
-\int_{\Sigma_C} d^2\sigma_{\mu\nu}(y)
\frac{1}{2}\epsilon^{\mu\nu\lambda}\partial_{\lambda }\frac{1}{|x-y|} 
\eeq
where the factor of $1/2$ appearing in the exponential is due to the
external quarks being in the fundamental representation; $\eta(x, C)$
is the solid angle subtended by the loop $C$ at the point $x$; and
$\Sigma_C$ is an arbitrary surface bounded by the counter $C$.  For
simplicity, $\Sigma_C$ will be assumed to lie in the $(x_1, x_2)$
plane. In the case of infinitely large Wilson loops, the solid angle
is $2\pi$ below the surface, and changes discontinuously across the
loop to $-2\pi$.  Physically $\eta(x, C)$ is the magnetic scalar
potential due to a dipole layer of unit strength on the sheet
$\Sigma_C$.  As expected, formula (\ref{wilson3}) can be expressed
exclusively in terms of the $3d$ Greens function which is the only
relevant element of the CG representation. Also note that, if the
fractional monopole charges of strength $1/N$ were present in the
system, an additional factor of $1/N$ would appear in the expression
for $\eta(x,C)$ in Eq.(\ref{wilson3}).  From Eq. (\ref{wilson3}),
$\Box\eta$ can be easily computed,
\beq
\label{source3}
\Box\eta=4\pi\Theta_{\Sigma_C}(x_1, x_2)\delta^{\prime}(x_3),~~~\eta=2\pi \Theta_{\Sigma_C}(x_1, x_2)
{\rm sgn}(x_3),
\eeq
where $\Theta_{\Sigma_C}(x_1, x_2)$ is unity within the surface
$\Sigma_C$ and zero outside.

As argued above, the transition from $3d$ to $4d$ is realized by the
replacement
\beq
\label{transition}
\frac{1}{4\pi|x-y|}\rightarrow\frac{1}{2\pi^2(x_{\mu}-y_{\mu})^2},~~~ 1\rightarrow \int dt ,
\eeq
where the numerical factors, $4\pi$ and $2\pi^2$, are the volumes of
the two- and three- dimensional unit spheres respectively, and appear
so that the Green functions are properly normalized, $ \Box G(x
y)=-\delta(x-y)$.  Furthermore, an integration over all time-slices
must be included to complete the transition to $4d$.  Carrying out the
replacements (\ref{transition}) in Eq. (\ref{wilson3}), the following
expression for $\eta(x, C)$ written in the covariant $4$ dimensional
form is arrived at,
\beq
\label{wilson4}
 \eta(x, C)=-\frac{1}{\pi} \int_{\Sigma_C} 
d^2\sigma_{\mu\nu}(y)\oint dx_{\sigma}\epsilon^{\mu\nu\lambda\sigma} 
 \partial_{\lambda } \frac{1}{(x-y)^2},
\eeq
here the integration over time-slices $\int dt$ for a static particle
was replaced by the integration along an arbitrary trajectory of a
particle, $\epsilon^{\mu\nu\lambda}dt\rightarrow
\epsilon^{\mu\nu\lambda\sigma} \oint dx_{\sigma}$.  In the static
limit, (\ref{wilson4}) is reduced to expression (\ref{wilson3}) as it
must by construction.  It is interesting to note that
Eq. (\ref{wilson4}) for $\eta $ (with appropriate normalization) is
formally similar to the expression for the linking number of a closed
oriented surface $ d^2\sigma_{\mu\nu}$ and closed oriented curve $
\oint dx_{\sigma}$ in four dimensions.  However, the surface $\Sigma_C$,
bounded by the Wilson loop, and a trajectory of the particle are not
closed manifolds; nevertheless $\eta$ can be interpreted as measuring
a solid angle, in addition $\frac{\eta}{2\pi}$ has an
integer-number-discontinuity when a particle crosses the surface
$\Sigma_C$ analogous to the $3d$ case (\ref{wilson3}), where for a
very large Wilson loop $\frac{\eta}{2\pi}=\pm 1$ just above and just
below the surface.

From Eq. (\ref{wilson4}), $\Box\eta$ can be calculated for the Wilson
loop in the $(x_1, x_2)$ plane with the following result,
\bea
\label{source4}
\Box\eta=\frac{1}{\pi}(2\pi^2)
\int_{\Sigma_C} 
d^2\sigma_{\mu\nu}(y) \oint dy_{\sigma}\epsilon^{\mu\nu\lambda\sigma} 
 \partial_{\lambda } \delta^4(x-y) \rightarrow
4\pi \Theta_{\Sigma}(x_1, x_2) \int dy_{\lambda}\epsilon^{\lambda\sigma}
\partial_{\sigma } \delta^2(x-y) , ~~ \lambda, \sigma=0, 3,
\eea
and is the final expression for the source of a Wilson loop in four
dimensions.  For the static case, the integration over $t$ can be
carried out and the above expression reduces to the $3d$ result found
in Eq. (\ref{source3}). For very large Wilson surfaces,
Eq. (\ref{source4}) suggests the following simplified expression for
$\eta$,
\beq
\label{eta}
\eta=  \Theta_{\Sigma}(x_1, x_2) \int dy_{\lambda}\epsilon^{\lambda\sigma}
\partial_{\sigma } \ln (x-y)^2 , ~~ \lambda, \sigma=0, 3.
\eeq
Indeed, applying the operator $\Box$ to Eq. (\ref{eta}), and taking
into account the relation $\Box \ln (x-y)^2=4\pi \delta^2(x-y)$,
reproduces Eq. (\ref{source4}).  Equation (\ref{source4}) explicitly
shows that $\eta/(2\pi)$ has a discontinuity whenever a particle
crosses the surface.  This is precisely the property of the Wilson
loop insertion.  Drawing attention to the behaviour in the SG picture,
Eq. (\ref{Z}), the Wilson loop insertion corresponds to the shift
$\phi\rightarrow\phi+\eta$ in the flavor singlet term of the potential
(\ref{Z}): $ E \cos \left( \frac{q}{p} \left(\phi+\eta - \theta + 2\pi
n \right) + 2\pi k \right) $.

The above discussions should have convinced the reader that although
the effective SG action for low energy QCD does not contain the
fundamental gauge degrees of freedom, it is possible to recover the
Wilson loop insertion operator by inserting the appropriate source
term. The two key points in making such a correspondence were:
\begin{enumerate}
\item 
$Q^{(0)}$ was identified with the physical magnetic charges;
consequently, the singlet combination of fields, $\phi$, was
identified with the physical magnetic scalar potential
\item 
Confinement is realized in this system through the condensation of the
magnetic charges $Q^{(0)}$ such that $\la \MM\ra\neq 0$.  In this case
the insertion of the Wilson loop in the monopole plasma leads to a
shift $\phi\rightarrow\phi+\eta$ of the corresponding magnetic scalar
potential as demonstrated above (\ref{bkfield}).
\end{enumerate}
One further point is that since the species $Q^{(0)}$ have fractional
charge of strength $1/N_c$, this additional factor should be
introduced in front of Eq. (\ref{wilson4}).

Even though the insertion of the Wilson loop operator in the CG
representation is now understood, the calculation of its VEV is much
more difficult in comparison with Polyakov's model \cite{Po77}.  The
fundamental difference is related to the fact that in Polyakov's case
the weak coupling regime is justified, and the classical action
dominates over the contributions from one loop fluctuations. In the
present case the only relevant dimensional parameter is the vacuum
energy, $E\sim \Lambda_{QCD}$, all other parameters are expressed in
terms of $E$ and are of the same order of magnitude; as such, there
are no small parameters in the problem.  Nevertheless, the VEV of the
Wilson loop can be estimated in the semi-classical approximation
analogous to Ref. \cite{Po77}. In this case for each given time-slice,
the problem is effectively a $3d$ problem described in \cite{Po77}
where the area law has been demonstrated $\la W \ra\sim \exp{( -S)}$
with calculable string tension.  In our $4d$ case, it is also expected
that $\la W \ra$ demonstrates the area law in agreement with our
earlier calculation of the magnetization, $\la \MM\ra\neq 0$; however,
an explicit calculation is still lacking.

To conclude this section, we note that a second important source
appears in this system due to the existence of domain wall solutions
to the equations of motion derived from (\ref{Z}). These domain walls
interpolate between vacuum states labeled by the parameter $k$, and
were discussed in some detail in \cite{FHZ}. Similar domain walls are
known to exist in supersymmetric models, and are reviewed quite
thoroughly in \cite{Shifman1}.  Recently, Witten conjectured
\cite{Wi98} that in the large $N$ limit the domain walls
connecting two vacua labeled by $k$ and $k+1$ appear to be object that
are not solitons, from the string viewpoint, but rather look like
$D$-branes on which the string should be able to end.  Such
$D$-brane-like domain walls have recently appeared in \cite{Wi98}
in the context of the AdS/CFT correspondence.  In fact, our CG picture
seems to indicate that such a phenomenon takes place in QCD as well.
Indeed, a distinguishable property of $D$-branes is that in large-$N$
limit its tension is $\sim N$. Domain walls described in\cite{FHZ}
have exactly this property\footnote{ To be more precise the wall
surface tension $\sigma\sim \Lambda_{QCD}^3 N^{1/2}+m_q\Lambda_{QCD}^2
N$, and in the chiral limit becomes $\sigma\sim \Lambda_{QCD}^3
N^{1/2}$; however, in the large-$N$ limit $\sigma\sim N$ as
expected.}. Additional support in favor of this identification comes
from our demonstration of the condensation of fractionally charged
$\sim a/N$ magnetic particles; which implies that the electric charges
in the system could also be fractional\footnote{ This can easily be
seen from Eq.(\ref{7}) by making a shift
$\theta\rightarrow\theta+2\pi$ (which does not change the physics) and
realizing that the system supports excitations with fractional
electric charges $Q\sim M+N$ along with fractional magnetic charges
$M$ even for $\theta=0 (mod 2\pi) $.}  and the domain walls support
excitations that carry electric charges $\sim 1/N$. Consequently, the
chromo-electric flux contained in the open string of the corresponding
(non-critical) string theory can end on the QCD $D$-brane.

\section{Conclusion}

The most important ``formal result'' of this paper is given by
Eqs. (\ref{CGexp}, \ref{CGaction}); which we claim is the new
representation of low energy QCD (\ref{Z}) that is appropriate for the
analysis of the vacuum structure.  The correspondence between these
two representations is discussed at length in the text. The most
important physical (as opposite to the ``formal'') result of the paper
is formulated as a conjecture described at the end of Section III in
which the charges, $Q^{(0)}_i$, from the statistical ensemble
(\ref{CGexp}, \ref{CGaction}) are identified with the
instanton-quarks\cite{Belavin}. If this conjecture is proven correct,
enormous progress in our understanding of QCD will have been made.
For example, it would fill the missing element of the well-developed
instanton picture\cite{Shuryak} to include the electric confinement of
quarks as a natural consequence of the {\it same } well-known
BPST-instantons\cite{BPST},\cite{Atiyah} which have been under
intensive study since the 70s.  Indeed, as discussed in the text, our
particles (which are the instanton-quarks, according to the
conjecture) gain the magnetic charges which condense, $\la \MM
\ra \neq 0$. Therefore, the standard 't Hooft-Mandelstam picture
\cite{Hooft} for the confinement would occur due to the instantons.

Various arguments which support this conjecture have been given in the
bulk of the text; however, one further reason why we believe this
conjecture could be correct will now be presented. Our CG
representation implies that the number of integrations $d^4x_i^{(0)}$
in (\ref{CGexp}) for each configuration is exactly equal to the
number, $M_0$, of fractionally charged particles, $Q^{(0)}\sim 1/p$,
for that configuration. Due to the fact that the total charge of the
configuration is integer, the number $M_0$ must be proportional to the
denominator of the charge, i.e $\sim p$; while, the fractional charge
$Q^{(0)}\sim 1/p$ in the CG representation appears due to the
$\theta/p$ dependence in the effective Lagrangian approach (\ref{Z}).
Consequently, as long as there is a $\theta/p$ dependence in the SG
representation, the number $M_0$ must be an integer multiple of $p$,
and the total dimensionality of integrations over $d^4x_i^{(0)}$ is
$4M_0= 4 p k$. This is exactly the measure in the multi-instanton
background with $p=C_2(G)$!  The quadratic Casimir operator
$C_{2}(G)$ appears here because in SUSY theories this number
determines the $\theta/C_2(G)$ dependence.

Indeed, it is well known since the work in \cite{SV} that the $\theta$
dependence of the gluino condensate $\la Tr
\lambda\lambda\ra_{\theta}$ in SYM with arbitrary gauge group $G$ is,
\beq
\label{gluino}
\la Tr \lambda\lambda\ra_{\theta , k}=
\la Tr \lambda\lambda\ra_{\theta=0}\exp\left(i~\frac{\theta+2\pi k}{C_2(G)}\right),
~~~ k=0, 1,..., ~ C_2(G)-1
\eeq
where $C_{2}(G)$, the quadratic Casimir operator, becomes equal to the
dual Coxeter number when the longest root vectors is normalized to
have length one.  In particular, $C_2(SU(N))=N$, $C_2(SO(N))=N-2$,
$C_2(Sp(2N))=N+1$.  Formula (\ref{gluino}) implies that for each given
$\theta$ there exist $C_2(G)$ degenerate vacuum states (which
corresponds to the Witten index) for which $\la Tr \lambda\lambda\ra$
differs by a phase factor of $\exp\left(i~\frac{2\pi
k}{c_2(G)}\right)$.  Then the $\theta$ evolution from $\theta=0$ to
$\theta=2\pi$ according to Eq. (\ref{gluino}) simply renumbers these
states in a cyclic way. The easiest way to understand this result is
(roughly speaking) to count the number of gluino zero modes in the one
instanton background (it is equal to $2C_2(G)$) such that the
correlation function $\la Tr \lambda\lambda, Tr \lambda\lambda
,...\ra_{\theta , k}\sim \exp(i\theta),$ including $C_2(G)$ insertions
of the operator $Tr \lambda\lambda$ is not zero.  Formula
(\ref{gluino}) then follows from the consideration of this correlation
function (for a complete analysis see the original paper \cite{SV}).
The number of bosonic zero modes is twice as much and equals to
$4c_2(G)$\cite{Instanton}.  The generalization of this result for an
arbitrary number, $k$, of instantons is also well known and is given
by the formula $4C_2(G)k$\cite{Instanton}.  These well known results
are reviewed here to place emphasis on the one-to- one correspondence
between $\theta/ C_2(G)$ dependence of the physical parameters and the
number of zero modes $4C_2(G)k$ in the $k$ instanton background.  But
as argued above, the total dimensionality of integrations over
$d^4x_i^{(0)}$ in our formula (\ref{CGexp}) exactly equals to this
number $4M_0= 4 C_2(G) k$, if the $\theta/ C_2(G)$ dependence in
non-supersymmetric gluodynamics remains the same as in SUSY
theories. Soft supersymmetry breaking analysis strongly suggests that
this is the case\cite{Shifman1}.  Therefore, our conjecture can be
reformulated in the following way: if $\theta/ C_2(G)$ dependence in
non-supersymmetric gluodynamics remains the same as in SYM, our
particles (\ref{CGexp}) could be nothing but the instanton-quarks
suspected long ago\cite{Belavin}.  The recent paper\cite{Mattis},
where it was demonstrated that in SYM the instanton-quarks carry
magnetic charges and saturate the gluino condensate, also supports our
picture.

As a final concluding remark, at the intuitive level there seems to be
a close relationship between our CG representation in terms of Abelian
monopoles, $Q_i^{(0)}$, and the ``Abelian projection'' approach
\cite{Diakonov,Polikarpov,Brower} as well as the ``periodic
instanton'' analysis \cite{vanbaal}. We also feel that there is a
close connection with the work of \cite{Fadeev} in the description of
infra-red QCD physics. Unfortunately, an explicit realization of these
correspondences is still missing at the moment; however, the search
for such mappings will be the focus of future studies.

\section{Acknowledgments}

This work is supported in part by the National Science and Engineering
Research Council of Canada.  S. J. would like to thank the
University of British Columbia for financial support through a
University Graduate Fellowship.

\section*{Appendix A: $\theta$-dependence in the massless limit  }

In this appendix the $\theta$-dependence of the partition function is
proven to vanish in the limit of zero quark masses.  This is an
absolutely trivial and well-known result of QCD. It is easily
understood from the effective Lagrangian approach (\ref{Z}), where the
$\theta$ dependence can be eliminated in the chiral limit by the shift
$\phi\rightarrow \phi-\theta$.  However, it is instructive to
understand this $\theta$ independence of $Z$ by direct analysis of the
statistical ensemble (\ref{CGaction}) with nonzero $\theta$ within the
CG representation.  It will be demonstrated that the $\theta$
independence is recovered in the chiral limit as a consequence of the
long range Coulomb interactions. Neglecting the interactions will
result in explicit $\theta$ dependence of the partition function,
which is clearly unwanted.  The first point to notice is that in this
limit only configuration which contain the species $Q^{(0)}$ are
allowed. This occurs since the mass dependence in the partition
function appears as $(m_a/2)^{M_a}$, for $a\ne0$, where $M_a$ is the
number of charges of species $Q^{(a)}$. Therefore in the massless
limit the partition function reduces to that of a single charge
species $Q^{(0)}$.

For $\theta\neq0$ it is convenient to label the configurations by the
number of positively, $n_1$, and number of negatively, $n_2$ charged
particles (note that for $\theta\ne0$ the $n_1$ and $n_2$ need not be
equal).  In this case the total charge of the
configuration\footnote{In this Appendix, without losing generality,
the charges $Q^{(0)}_i$ are rescaled to be $\pm 1$.} $Q^{(0)}_T=
\sum_{i=1}^{M_0}Q^{(0)}_i=n_1-n_2$, while the total number of
pseudo-particles $M_0=n_1+n_2$ is fixed and the corresponding
contribution, $Z^{(M_0,Q)}$, to the partition function
(\ref{CGaction}) is,
\bea
Z^{(M_0,Q)}(\theta)&\equiv& \exp(-iQ\theta)\exp(-F^{(M_0,Q)}) \nn\\
\label{a1}
&=& \exp(-iQ\theta)
\frac{(E/2)^{M_0}}{n_1! n_2!}\prod_{i=1}^{M_0} d^4x_i~
\exp\left\{-\sum_{i\neq j}^{M_0}\frac{Q_iQ_j}{|x_i-x_j|^2}\right\}
\eea
where, without losing generality, only the first branch with $k=0$
appears here; the superscript $(0)$ on $Q$ has been removed since
only a single charge species is present; $E$ is the fugacity of the
Coulomb gas; the combinatorical factor $\frac{M_0!}{n_1! n_2!}$ has
been introduced for the correct counting of $n_1$ identical particles
with charge $(+)$ and $n_2$ identical particle with charge $(-)$; and,
\beq
Q\equiv Q^{(0)}_T  = n_1-n_2,\ \ \ M_0=n_1+n_2, \ \ \ Q_i=\pm 1
,\ \ \ i=1,...M_0,
\eeq
For the special case of vanishing $\theta$, the neutrality condition
would require $n_1=n_2$ in the thermo-dynamical limit. Now, if the
Coulomb interaction could be ignored, then (\ref{a1}) yields a free
energy $F(\theta)\simeq - \log Z(\theta)$ which is a $\theta$-{\it
dependent} expression.  Indeed, the partition function for a
noninteracting gas with a fixed total charge $Q$,
\begin{equation}
\label{a2}
Z^Q(\theta)\equiv \exp(-iQ\theta)\exp(-F^Q)=
\sum_{M_0}Z^{(M_0,Q)}(\theta)
\rightarrow \int dM_0~Z^{(M_0,Q)}(\theta)
\end{equation}
can be obtained by the steepest descent method with respect to $M_0$.
If the size of the system, $L$, is large, the noninteracting case
reduces to,
\bea
\exp(-F^Q)&=&\int dM_0~\exp\left(M_0\ln(L^4E)\right)\times 
\exp \left\{\frac{M_0+Q}{2}\left(\ln\frac{M_0+Q}{2}-1\right)
-\frac{M_0-Q}{2}\left(\ln\frac{M_0-Q}{2}-1\right)\right\} \nn\\
\label{a3} 
&\sim& \exp(\overline{M}_0)~\exp\left\{-\frac{Q^2}{2\overline{M}_0}\right\}
\eea
where the formula $\ln(k!)\sim k\ln k-k$ has been used, and
$\overline{M}_0$ is the saddle point determined from the equation,
\begin{equation}
\label{a4}
\ln(L^4E)=\frac{1}{2}\ln\left\{\frac{{\overline{M}_0}^2-Q^2}{4}\right\}, \qquad
\frac{Q}{M_0}\ll 1, \qquad \overline{M}_0\simeq 2L^4E
\end{equation}

In this approximation of noninteracting particles, the partition
function is evaluates to,
\beq
Z(\theta)\equiv\exp(-F(\theta))\sim\exp(2L^4E)\int dQ
\exp(-i\theta Q)\exp\left\{-\frac{Q^2}{4E L^4}\right\} 
\sim \exp\left\{2E L^4\left(1-\frac{\theta^2}{2}\right)\right\}
\label{a5}
\eeq
Consequently, the free energy is given by,
\beq
F(\theta)\simeq -2E L^4 \left(1-\frac{\theta^2}{2}\right), 
\qquad\theta \ll 1 
\eeq
The evaluation of $Z$ appearing in (\ref{a5}) reveals that the
essential terms are those which contain the charges,
\begin{equation}
\label{}
|Q|\sim EL^4\theta, \ \ \ \ \ \ \ \ \ \frac{|Q|}{M_0}\sim\theta
\end{equation}
and the approximation $Q/M_0 \ll 1$ can be justified for small $\theta
\ll 1$ which will be assumed henceforth.  Therefore, as
expected, the free energy $F(\theta)$ for the gas of the
noninteracting particles explicitly depends on $\theta$.

The discussion now returns to the full long-range interacting Coulomb
gas (\ref{a1}).  This problem is in all respects very similar to
Polyakov's model\cite{Po77}.  It is known from \cite{Vergeles} that
the Coulomb interactions in a plasma cannot be ignored in this
case. Hence, formula (\ref{a5}) is certainly misleading. In order to
derive some estimation for the interacting case, once again the system
is placed into a box of size $L$. In this case, excessive charge is
deposited on the walls of the box, and the free energy is that of a
neutral Debye plasma plus the Coulomb energy, similar to what happens
in the well-understood $QED_3$ case \cite{Vergeles},
\bea
\label{a6}
\exp(-F^Q)\sim\!\!\int\!\! dM_0 ~\exp\left(M_0  \ln(L^4E)\right)\times  
\exp\left\{\frac{M_0+Q}{2}\left(\ln\frac{M_0+Q}{2}-1\right)
-\frac{M_0-Q}{2}\left(\ln\frac{M_0-Q}{2}-1\right)-\frac{Q^2}{L^2}\right\}
\eea
The partition function with a fixed total charge $Q$ can now be
estimated as in (\ref{a3},\ref{a4}), with the only difference occurring
in the last term which is due to the Coulomb interaction.  This term
plays a key role in the following integration over all charges $\int
dQ$,
\begin{equation}
\label{a7}
Z(\theta)\equiv\exp(-F(\theta))\sim\exp(2 L^4E)\int dQ~
\exp(-i\theta Q)~
\exp\left\{-\frac{Q^2}{4E L^4}\right\}~
\exp\left\{-\frac{Q^2}{L^2}\right\}
\end{equation}
Taking into account the extra factor $\exp(-\frac{Q^2}{L^2})$, the
charges of the essential configurations behave as,
\begin{equation}
\label{a8}
|Q|\sim L^2\theta ,\ \ \ \ \ \ \frac{Q}{M_0}\sim\frac{\theta}{ L^2}
\rightarrow 0~~ at~~ L\rightarrow\infty 
\end{equation}
consequently,
\begin{equation}
\label{a9}
F(\theta)\sim -2E L^4+L^2\theta^2
=-2E L^4\left\{1-\frac{\theta^2} {2 L^2}\right\}
\end{equation}
Thus, as $L\rightarrow \infty$, the free energy $F(\theta)$ in $QCD_4$
in the chiral limit does not depend on $\theta$, which is precisely
how it should behave and is exactly what is expected from the
Sine-Gordon representation (\ref{Z}) of the partition function.  It
should be emphasized once more that this result is a direct
consequence of the strong Coulomb interactions in the system.

\section*{Appendix B: Large gauge transformations and
monopole charges} 

In this Appendix the large gauge transformations generated by the
operator $N$, as discussed in the text (\ref{7}), will be elaborated
on. It will be demonstrated that such a transformation is equivalent
to the identity operator. In the following analysis it will be
assumed, along the lines of `t Hooft \cite{Hooft1}, that formula
(\ref{7}) has much more general applicability and is not necessarily
constrained to the weak coupling limit of the Georgi-Glashow model
where it was originally derived.  Due to knowledge of the
$\theta$-dependence in the effective low energy potential on both
sides of Eq. (\ref{7}), the magnetic properties of the relevant fields
describing the large distance physics will be identified.

The strategy is to begin with the QCD effective anomalous potential
before the glue-ball degrees of freedom are integrated out. Next, a
large gauge transformation will be performed on this effective
potential. It is clear that this operation is equivalent to applying
the identity operator; however, in the course of the following
calculations, the magnetic and electric properties of the fields
making up the effective potential will be identified..
 
Recall that the effective potential is defined as the Legendre
transform of the generating functional for zero momentum correlation
functions of the marginal operators $ G_{\mu \nu} \tilde{G}_{\mu \nu}
$, $ G_{\mu \nu} G_{\mu \nu} $ and $m\bar{\Psi}^{i} \Psi^{i}$, see
\cite{HZ},\cite{FHZ} for details.  It is a function of the effective zero
momentum fields $ h , \bar{h}$ which describe the VEV's of the
composite complex fields $ H ,\bar{H} $, 
$$
\int dx \, h = \left\la \int dx \, H \right\ra \quad, \quad
\int dx \, \bar{h} =  \left\la \int dx \, \bar{H} \right\ra 
$$
where,
\beq
\label{b2}
H  = \frac{1}{2} \left( \frac{ \beta(\alpha_s)}{4 \alpha_s}
G^2  + i \, \frac{2p}{q} \frac{ \alpha_s}{4 \pi} 
G \tilde{G} \right) \; , \; \bar{H}   
= \frac{1}{2} \left( \frac{ \beta(\alpha_s)}{4 \alpha_s}
G^2  - i \, \frac{2p}{q} \frac{ \alpha_s}{4 \pi} 
G \tilde{G} \right)  \; .
\eeq 
The potential is also function of the unitary matrix $ U_{ij} $
corresponding to the $ \gmf $ phases of the chiral condensate: $ \la
\bar{\Psi}_{L}^{i} \Psi_{R}^{j} \ra = - | \la \bar{\Psi}_{L} \Psi_{R}
\ra | \, U_{ij} $.  As explained in Section 2 the corresponding
Lagrangian reproduces the chiral and conformal anomalies and also
corresponds to the following behavior of the $(2k)^{\rm th}$
derivatives of the vacuum energy in pure gluodynamics,
\beq
\label{b3}
 \left. \frac{ \partial^{2k} E_{vac}(\theta)}{ \partial \, \theta^{2k}}
\right|_{\theta=0}
 \sim \int \prod_{i=1}^{2k} d^4x_i \la Q(x_1)...Q(x_{2k})\ra
\sim \left(\frac{i}{N_c}\right)^{2k}, 
\eeq
where $Q\sim G_{\mu\nu} {\widetilde G}_{\mu\nu}$. This is a
consequence of the solution of the $U(1)$ problem when Veneziano
ghosts saturate all relevant correlation functions (\ref{b3})
\cite{Venez}.

Our starting point is then the effective QCD potential $W(h, U)$ which
in Minkowski space takes the form\cite{HZ},
\bea
\label{b4}
e^{- i V W(h, U) } &=& \sum_{n = - \infty}^{
 + \infty} \sum_{k=0}^{q-1} \exp \left\{ - \frac{i V}{4}
\left( h \, \log \, \frac{h}{2 e E} + 
\bar{h} \, \log \, \frac{ \bar{h}}{
2 e E } \right) \right. \nonumber \\ 
&&\hspace{25mm}+ \left. i \pi V \left( k + \frac{q}{p} \,  
\frac{ \theta - i \log Det \, U + 2 
\pi n}{ 2 \pi} \right) \frac{h - \bar{h}}{
2 i} + \frac{i}{2}  V \,{\rm Tr}( M U + h.c.) \right\} \;  ,
\eea
where $ M = {\rm diag} (m_{i} | \la \bar{\Psi}^{i} \Psi^{i} \ra | )$ ,
$V$ is the volume of the system and the complex fields $ h , \bar{h} $
are defined as in Eq.(\ref{b2}).
   
The anomalous effective potential (\ref{b4}) contains both the light
chiral fields $ U $ and heavy ``glue-ball" fields $ h \, , \,
\bar{h}$, and is therefore not an effective potential in the Wilsonian
sense. On the other hand, only the light degrees of freedom, described
by the fields $ U $ are relevant for the low energy physics. An
effective potential for the $ U \, , \, \bar{U} $ fields can be
obtained by integrating out the $ h \, , \, \bar{h} $ fields in
Eq.(\ref{b4}). In SUSY models the transition from the effective
potential for the $ U $, $ h $ fields to the effective potential for
the $ U $ fields, by integrating out the $ h \, , \, \bar{h} $ fields,
is analogous to the transition from effective Lagrangian \cite{TVY}
for SQCD to the Affleck-Dine-Seiberg \cite{ADS} low energy effective
Lagrangian.  The corresponding derivation for QCD was described
earlier in \cite{HZ,FHZ}. The goal here is to study the properties of
the various fields from (\ref{b4} ) under large gauge
transformations. In order to gain an understanding of what changes
this large gauge transformation produces, the complex $h, \bar{h}$-
fields are parameterized by introducing two real fields $ \rho$ and $\omega $,
\beq
\label{b5}
h = 2 E \, e^{ \rho + i \omega} \; \; , \; \; 
\bar{h} = 2 E \, e^{ \rho - i \omega } \; .
\eeq
The summation over the integers $ n $ in Eq.(\ref{b4}) then enforces the
quantization rule due to the Poisson formula,
\beq
\label{b6} 
\sum_{n} \exp \left(
2 \pi i n  \, \frac{q}{p} \, V \, \frac{h - \bar{h}}{ 4 i} 
\right) = \sum_{m} \delta \left( \frac{q}{p}
\, V E e^{ \rho} \, \sin \omega - m \right) \; ,
\eeq 
This reflects quantization of the topological charge $Q= \left(
\frac{q}{p} \, V E e^{ \rho} \, \sin \omega \right)$ which must be an
integer, $m$, in the original theory. Applying a large gauge
transformation, $T={\rm e}^{i 2\pi N}$, then implies that $m$ in
Eq. (\ref{b6}) undergoes the shift $m\rightarrow m- 1$. Indeed, by
definition $$ Z\sim \sum_{n}\sum_{m}\la n | e^{-S-i\theta Q}|m\ra
,\qquad Q=m-n, $$ where the $|\theta\ra$ state is the eigenstate of
the large gauge transformation operator, $T$, with eigenvalue
$e^{-i\theta}$.  Consequently, when a large gauge transformation is
performed the topological charge of the configuration becomes shifted
$Q\rightarrow Q+1$. This corresponds to the replacement $m\rightarrow
m- 1$ in Eq. (\ref{b6}). It is quite clear that such a replacement
does not change any physical content of the theory due to the
summation over $m$ in Eq. (\ref{b6}). Therefore, the large gauge
transformation is equivalent to the identity operation as expected.
However, the transformation properties of the individual fields under
such an operation is what needs to be understood.  Hence, proceeding
further, and using (\ref{b6}), Eq.(\ref{b4}) (with the mass term
omitted) is placed in the form,
\beq
\label{b7}
e^{ - i V W} = \sum_{m = - \infty
}^{ + \infty} \sum_{k=0}^{q-1}
 \delta\left( V E \, \frac{q}{p} \, e^{ \rho} 
\sin \omega - m\right) \, \exp \left\{ - i V E \, e^{ \rho} ( \rho -1)
\cos \omega  + i m \left( \bar{\theta}_k +  \frac{p}{q} \, \omega 
\right) \right\} 
\eeq
where,
\beq
\label{b8}
\bar{\theta}_k \equiv \theta - i \log Det \, U 
+ 2 \pi \, \frac{p}{q} \, k \; .
\eeq
To resolve the constraint imposed by the presence of the $
\delta$-function in Eqs.(\ref{b6}),(\ref{b7}), a Lagrange multiplier
field, $\Phi$, is introduced,
\beq
\label{b9}
 \delta( V E \, \frac{q}{p} \, e^{ \rho}
\sin \omega - m) \propto \int D \, \Phi \; \exp \left( i \Phi 
V E \, e^{ \rho } \sin \omega - i \Phi 
\, \frac{p}{q} \, m \right) \; 
\eeq
Wick rotating to Euclidean space by the substitution $ i V \rightarrow
V$, from Eqs.(\ref{b6}),(\ref{b9}) one obtains,
\bea
\label{b10}
W( U , \rho, \omega, \Phi) &=& - \frac{1}{V} \log \left\{ \sum_{m =
- \infty}^{+ \infty} \sum_{k=0}^{ q-1} \exp 
\left[
- V E e^{ \rho} \left\{ ( \rho - 1) \cos \omega - \Phi
\sin \omega \right\} 
  \right. \right.\nonumber \\ 
&& \hspace{10mm}+  \left. \left.
i m  \left( \theta 
- i \log Det \, U + 2 \pi k \, \frac{p}{q} + \frac{p}{q} \, \omega - 
\frac{p}{q} \, \Phi \right) -  \varepsilon \, 
\frac{m^2}{ VE} \right] 
\right\} \; .
\eea
The last term has been introduced to regularize the infinite sum over
the integers $ m $, and the limit $ \varepsilon \rightarrow 0 $ will
be taken at the end, but before taking the thermodynamic limit $ V
\rightarrow \infty $. Note that Eq.(\ref{b10}) satisfies the condition
$ W (\omega + 2 \pi) = W (\omega) $ as it should since $
\omega $ is an angular variable (\ref{b5}). Note also that Eq. (\ref{b10}) 
has an explicit $2\pi$ periodicity in $\theta$. As mentioned above,
the large gauge transformations lead to the shift $m\rightarrow m-1$;
therefore, the extra term which appears as a result of the action of
the operator $T$ is the phase proportional to $m$ in Eq. (\ref{b10}),
\bea
\label{b11}
T \rightarrow  e^{ -i   \left( \theta 
- i \log Det \, U + 2 \pi k \, \frac{p}{q} + \frac{p}{q} \, \omega - 
\frac{p}{q} \, \Phi \right)}.
\eea 
However, this operation must be the identity; consequently, this phase
must be unity at the very end of the calculations. Rather than
imposing this condition immediately, the transformation can now be
used to understand how this identity is realized in terms of the
contributions from different fields and/or vacuum charges at infinity.
To answer this question the VEV's for each field entering
Eq. (\ref{b11}) must be calculated in the thermodynamic limit.  The
thermodynamic limit $ V \rightarrow \infty $ of the potential
(\ref{b10}) can be obtained by making use of the Jacobi identities,
\beq
\label{b12}
 \theta_3(\nu, x)=\frac{1}{\sqrt{\pi x}}\sum_{k=-\infty}^{+\infty}
\exp \left\{ -\frac{(\nu+k)^2}{x} \right\} =
  \sum_{l=-\infty}^{+\infty}
 \exp [ -l^2\pi^2x +2il\nu\pi ]
\eeq
which allows Eq.(\ref{b10}) to be rewritten as,
\bea
\label{b13}
W (U,  \rho, \omega, \Phi) &=& - \frac{1}{V} \log \left\{ \sum_{n =
- \infty}^{ + \infty} \sum_{k=0}^{ q-1} \exp 
\left[
- V E e^{ \rho} \left\{ ( \rho - 1) \cos \omega -
 \Phi \sin \omega \right\} 
  \right. \right. \nonumber \\ 
&&\hspace{15mm}-  \left. \left.
\frac{VE}{ 4 \varepsilon} 
\left( \theta - i \log Det \, U +  2 \pi k \, \frac{p}{q} + 
\frac{p}{q} \, 
\omega - 
\frac{p}{q} \, \Phi - 2 \pi n \right)^2  \; \right] \right\} \; ,
\eea
Here, an irrelevant overall infinite factor $ \sim \varepsilon^{-1/2}$
has been ignored.  Eq.(\ref{b13}) is the final expression for the
effective potential $ W $, which is suitable for analysis of the
thermodynamic limit.  Such an analysis was carried out in \cite{FHZ}
where it was shown that the coordinate-independent solutions,
corresponding to the vacuum states, take the following form,
\beq
\label{b14}
\la \omega \ra_l = - \frac{q}{p} \, 
( \theta  - i \log Det \, U ) + 2\pi \frac{q}{ p}r
  - 2 \pi k \quad, \quad
 \la \rho \ra = 0 \quad  , \quad  \la \Phi \ra = 0 \quad , 
\; \qquad r = 0, \pm 1, \ldots , \qquad k = 0,1, \ldots , q-1 \;
\eeq
The combination
$2\pi \frac{q}{ p}r
  - 2 \pi k $
  which enters this equation can be represented as 
 \beq
\label{b15}
2\pi \frac{q}{ p}r
  - 2 \pi k  =   \frac{2\pi}{ p}
\, l - 2 \pi r'~~~~~~~r, r' = 0, \pm 1, \ldots ,~~~~~~~ l= 0,1, \ldots , p-1 \;
\eeq 
Eqs.(\ref{b14},\ref{b15}) show that there are $ p\sim N_c $ physically
distinct solutions of the equation of motion for the $ \omega $ field,
while the series over the integers $ r' $ in Eq.(\ref{b15}) simply
reflects the angular character of the $ \omega $ variable, and is
therefore irrelevant.  Substituting Eq.(\ref{b14}) into
Eq.(\ref{b11}), the phase which appears due to the insertion of the
large gauge transformation operator is seen to be identically zero
(mod $2\pi n$) as expected. What is important now is the possibility
to interpret this result in terms of the specific superposition of the
following operations: 
\begin{enumerate}
\item
$e^{- \log Det \, U}=e^{i\phi(x)}$. The local field $\phi$ has
already been identified with the magnetic scalar potential. Therefore,
a nonzero vacuum expectation value for $\la e^{i\phi(x)}\ra $ implies
the monopole condensation and nonzero magnetization as discussed in
section IV.
\item
$e^{-i \frac{p}{q} \omega}$. The local field $\omega$ was defined as
the phase of the fields $h, \bar{h}$ (\ref{b5}) which describe the
vacuum expectation value of the composite fields $ H ,\bar{H} $
(\ref{b2}). A nonzero vacuum expectation value for this operator $\la
e^{-i\frac{p}{q} \omega}\ra $ also implies that the condensation of
monopoles is related somehow to the glue-ball degrees of freedom $ H
,\bar{H} $ (\ref{b2}); however, a precise statement cannot be made
because the CG representation (\ref{CGaction}) was derived from SG
effective Lagrangian (\ref{Z}) after the integrating out all glue-ball
fields $h, \bar{h}$.
\item
$e^{-i (\theta + 2 \pi k \, \frac{p}{q})}$. This operator is not
related to a local field, but rather, is related to the background
charge at infinity just like in the well-known example of the $2d$
Schwinger model.
\end{enumerate}

To conclude: In section III it was demonstrated that the charges
$Q^{(0)}$ from our system (\ref{CGaction}) can be identified with the
magnetic charges (\ref{9}), and the field $\phi$ can be identified
with the magnetic potential (\ref{Q0insert}).  In this Appendix the
next step in analysis of the CG-SG correspondence has been carried
out.  The large gauge transformation (\ref{7}) was performed on the
potential and it was demonstrated that the expected identity is
realized as the superposition (combination) of three different
non-zero contributions related to the condensation of: 1) the quark
fields, 2) the gluon fields and 3) the background charges at infinity
as explained above.

 \end{document}